\documentclass[a4paper,fleqn]{cas-dc}
\usepackage[numbers,sort&compress]{natbib}
\usepackage[T1]{fontenc}
\usepackage[english]{babel}
\usepackage{graphicx}
\usepackage{amsmath}
\usepackage{booktabs}
\usepackage{array}
\usepackage[version=4]{mhchem}
\usepackage{hyperref}
\usepackage{epstopdf}
\newcommand{\missinggraphic}[2][]{\IfFileExists{#2}{\includegraphics[#1]{#2}}{\fbox{\begin{minipage}[c][3cm][c]{0.85\linewidth}\centering\footnotesize Figure not provided: \texttt{\detokenize{#2}}\end{minipage}}}}

\begin{document}
\shorttitle{Janus Orientation-Controlled Charge Transfer in BX/WSTe}
\shortauthors{Chaoui et al.}
\title[mode=title]{Janus Orientation-Controlled Charge Transfer and Absorptance-Weighted Solar-to-Hydrogen Efficiency in BX/WSTe Heterostructures}

\author[unamur]{Khawla Chaoui}
\cormark[1]
\ead{khawla.chaoui@unamur.be}
\author[guelma]{Warda Elaggoune}
\author[unamur,um6p]{Mohamed Achehboune}
\author[unamur]{Luc Henrard}
\cortext[cor1]{Corresponding author.}
\affiliation[unamur]{organization={Department of Physics, Namur Institute of Structured Matter, University of Namur},
 addressline={Rue de Bruxelles 61}, city={Namur}, postcode={5000}, country={Belgium}}
\affiliation[guelma]{organization={Laboratoire de Physique des Mat\'eriaux (L2PM), Universit\'e 8 Mai 1945},
 addressline={BP 401}, city={Guelma}, country={Algeria}}
\affiliation[um6p]{organization={College of Physical Sciences and Engineering, University Mohammed VI Polytechnic},
 city={Ben Guerir}, postcode={43150}, country={Morocco}}

\begin{abstract}
Solar-driven water splitting requires efficient light harvesting, favorable redox energetics, and interfacial charge separation. Here, we investigate Janus BX/WSTe (X = P, As) van der Waals heterostructures using first-principles calculations, focusing on the effects of Janus orientation on interfacial charge transfer and solar-to-hydrogen (STH) efficiency.
Reversing the orientation of WSTe reverses the direction of interfacial electron transfer and modifies the electronic properties of the interface. BP/TeWS and BAs/SWTe exhibit characteristics consistent with possible direct Z-scheme pathways, whereas BAs/TeWS displays conventional type-II characteristics. Band-edge analysis identifies thermodynamically favorable conditions for overall water splitting at pH below 6.6 for BP/TeWS and between 3.7 and 8.1 for BAs/TeWS, while BAs/SWTe supports only hydrogen evolution under conventional aqueous conditions.
By incorporating the calculated energy-dependent in-plane absorptance into AM1.5G solar-spectrum integrals, the theoretical STH efficiencies reach approximately 2.91\% for BP/TeWS and 3.20\% for BAs/TeWS, representing a reduction of approximately 91–92\% relative to surface-potential-corrected full-absorption estimates. These findings demonstrate the importance of Janus orientation in controlling interfacial properties and of realistic optical absorptance in evaluating two-dimensional photocatalysts.
\end{abstract}
\maketitle

\section{Introduction}\label{introd}

Solar-driven photocatalytic water splitting offers a promising route for converting abundant solar energy into dihydrogen, a clean energy carrier that can be produced without direct carbon emissions \cite{nishioka2023photocatalytic,singla2021photocatalytic}. Since the pioneering work of Fujishima and Honda in 1972, who demonstrated photoelectrochemical water splitting using TiO$_2$ \cite{fujishima1972electrochemical}, considerable efforts have been devoted to developing efficient photocatalytic materials. Despite its chemical robustness, the wide band gap of TiO$_2$ ($\sim$3.2 eV) limits its optical response mainly to the ultraviolet region \cite{guo2019fundamentals}, although strategies to enhance its light harvesting through structural engineering have been proposed \cite{habli2025computational}. Narrower-band-gap semiconductors, such as CdS and Cu$_2$O, have subsequently attracted attention for visible-light-driven photocatalysis \cite{baniasadi2013hybrid,he2015novel,kwon2015shape,grozdanov1994electroless}. However, these materials still face challenges related to rapid charge-carrier recombination and photocorrosion \cite{CdS:review2023}. These limitations highlight that a suitable band gap alone is insufficient for efficient overall water splitting. An effective photocatalyst must simultaneously ensure efficient solar-light harvesting, effective charge separation, suitable redox potentials for both the hydrogen evolution reaction (HER) and oxygen evolution reaction (OER), and sufficient stability under operating conditions \cite{li2022challenges,bie2022challenges,maeda2010photocatalytic,maeda2011photocatalytic}.

Two-dimensional (2D) materials have emerged as promising candidates for addressing several of these requirements because of their large surface-to-volume ratio, short carrier-transport distances, and highly tunable electronic structures \cite{novoselov2004electric,yuan2023first,chhowalla2013chemistry,novoselov20162d,Elaggoune2025_BX}. Their reduced dimensionality provides abundant exposed surface sites and allows their electronic and optical properties to be tailored through composition, strain, external fields, and interfacial engineering. Nevertheless, an isolated 2D semiconductor rarely satisfies all the energetic and kinetic requirements for overall water splitting. In particular, strong optical absorption does not necessarily ensure efficient charge separation, while favorable band-edge positions may be accompanied by rapid electron--hole recombination \cite{hisatomi2014recent,kang2021electronic,jun2023two}. These competing requirements have motivated the development of 2D heterostructures, in which complementary materials can be combined to facilitate spatial charge separation while retaining desirable electronic and optical characteristics \cite{novoselov20162d,kudo2009heterogeneous,low2017heterojunction,zhang2012recent}. Among these systems, vertically stacked van der Waals heterostructures (vdWHs) provide a particularly versatile platform because their constituent layers can be assembled without the stringent lattice-matching constraints imposed by conventional covalent interfaces \cite{gao2019zno,niu2017arsenene,ran2021significantly}.

Depending on the relative positions of the band edges of the constituent semiconductors, van der Waals heterostructures can exhibit different band alignments. In type-I (straddling-gap) heterostructures, both the conduction-band minimum (CBM) and valence-band maximum (VBM) are localized within the same component, favoring carrier confinement. In contrast, type-II (staggered-gap) heterostructures exhibit CBM and VBM states localized on different components, promoting spatial separation of photogenerated electrons and holes \cite{li2018type,zhou2020type,mao2021hierarchically}. Although conventional type-II charge transfer can facilitate carrier separation, it generally reduces the available redox ability because the surviving electrons and holes occupy the lower-energy conduction-band and higher-energy valence-band states, respectively \cite{huang2017switching}.

A direct Z-scheme pathway offers an alternative mechanism that can preserve stronger redox capabilities while promoting charge separation. In such systems, differences in the Fermi levels of the constituent materials can induce interfacial charge redistribution upon contact, generating a built-in electric field (IEF). Under illumination, this field can favor the interfacial recombination of photogenerated electrons and holes with weaker redox abilities, leaving the more strongly reducing electrons and oxidizing holes on opposite components \cite{low2017review,xu2018direct}. Consequently, although conventional type-II and direct Z-scheme systems may exhibit similar staggered band arrangements, they differ in their proposed carrier-transfer pathways and the redox capabilities of the remaining photogenerated carriers. Direct Z-scheme mechanisms have been investigated in various photocatalytic systems, including CdS/TiO$_2$, g-C$_3$N$_4$/TiO$_2$, aza-CMP/C$_2$N, black-phosphorus-based heterostructures, and C$_3$B/C$_3$N \cite{meng2017direct,jo2015influence,wang2018van,hu2019z,niu2020rational}, as well as in more recent two-dimensional van der Waals heterostructures \cite{Chaoui2025_InNPtSSe,Chaoui2024_BAsGeC}. The possibility of controlling the interfacial electronic conditions associated with conventional type-II or direct Z-scheme pathways is therefore particularly attractive for rational photocatalyst design.

Janus 2D materials introduce an additional degree of control because their two surfaces are chemically inequivalent. Replacing one chalcogen plane of a transition-metal dichalcogenide with a different chalcogen breaks the out-of-plane mirror symmetry and generates an intrinsic electrostatic asymmetry. Consequently, the two surfaces can exhibit different work functions, leading to orientation-dependent interfacial charge transfer. The direction and magnitude of this charge transfer can therefore depend on which Janus surface is brought into contact with the adjacent layer \cite{wang2021constructing,tang2023engineering,alrebdi2022situ}. Since the experimental realization of monolayer Janus MoSSe \cite{zhang2017janus,hernandez2022janus}, Janus transition-metal dichalcogenides have attracted considerable interest for optoelectronic and photocatalytic applications \cite{Chaoui2024_GeSnXY,Chaoui2025_XGaPbP}. First-principles investigations of MXY monolayers and Janus-based heterostructures have demonstrated substantial tunability of band alignment, interfacial electric fields, and photocatalytic properties \cite{riis2019classifying,idrees2020van,wang2022two}. In particular, WSTe represents an interesting building block because its pronounced out-of-plane asymmetry can generate markedly different interfacial electrostatic environments depending on whether the S or Te surface forms the interface \cite{idrees2019optoelectronic}.

Beyond controlling interfacial electronic properties, accurately evaluating the solar-to-hydrogen (STH) conversion efficiency remains an important challenge in theoretical photocatalyst screening. Theoretical STH models rely on material-dependent parameters obtained from first-principles calculations, other modeling approaches, or experimental measurements. In many first-principles screening studies, the light-harvesting contribution is estimated by assuming that all incident photons with energies above the band gap are effectively absorbed \cite{fu2018intrinsic}. While this idealized approximation is useful for estimating intrinsic energetic limits, it can substantially overestimate the light-harvesting capability of atomically thin photocatalysts, for which even a large absorption coefficient does not necessarily imply strong absorption of incident radiation. For a single 2D layer, the relevant optical quantity is the absorptance, which represents the fraction of incident optical power absorbed at each photon energy. Explicitly incorporating the energy-dependent 2D absorptance into the AM1.5G solar-spectrum integrals therefore provides a more realistic evaluation of the light-harvesting contribution to the theoretical STH efficiency. Moreover, the framework proposed by Fu et al. accounts for the intrinsic out-of-plane potential difference in the evaluation of STH efficiency \cite{fu2018intrinsic}, making it particularly relevant to polar Janus-based photocatalysts. Combining orientation-dependent interfacial electrostatics with an absorptance-weighted efficiency evaluation thus provides a useful framework for assessing the photocatalytic potential of Janus-based heterostructures.

In this work, we investigate BX/WSTe ($X$ = P, As) Janus van der Waals heterostructures using first-principles calculations, focusing on the interplay between Janus orientation, interfacial charge transfer, and photocatalytic performance. Six stacking configurations and both possible contact orientations of WSTe are considered to examine how its intrinsic electrostatic asymmetry influences structural stability, band alignment, interfacial charge redistribution, and the built-in electric field. Our results reveal that reversing the orientation of WSTe reverses the direction of interfacial electron transfer, giving rise to distinct charge-transfer characteristics. BP/TeWS and BAs/SWTe exhibit electronic and interfacial features consistent with direct Z-scheme pathways, whereas BAs/TeWS displays characteristics of conventional type-II charge transfer. Band-edge analysis identifies BP/TeWS and BAs/TeWS as thermodynamically suitable for overall water splitting within different pH ranges: BP/TeWS satisfies the required redox conditions for pH below approximately 6.6, while BAs/TeWS is suitable within $3.7 < \mathrm{pH} < 8.1$. In contrast, BAs/SWTe supports only the hydrogen-evolution half-reaction within the conventional aqueous pH range. Furthermore, we calculate the energy-dependent in-plane absorptance of the heterostructures from their two-dimensional optical response and explicitly incorporate it into the AM1.5G solar-spectrum integrals. The resulting absorptance-weighted theoretical STH efficiencies range from 2.87\% to 2.91\% for BP/TeWS and from 3.17\% to 3.20\% for BAs/TeWS, depending on pH. These values are substantially lower than those obtained under the ideal full-absorption assumption, emphasizing the importance of accounting for the actual optical absorptance of atomically thin materials. By combining orientation-dependent interfacial electrostatics with an absorptance-weighted STH evaluation, this study provides insights into the assessment and design of Janus-based two-dimensional photocatalysts for solar-driven water splitting.

\section{Computational details}
First-principles calculations based on density functional theory (DFT) were performed using the Vienna Ab initio Simulation Package (VASP) \cite{vasp1:kresse1993ab,vasp2:kresse1996efficient}. The interaction between ions and electrons was described using the projector-augmented-wave (PAW) method \cite{PAW:blochl1994projector}, while the exchange--correlation effects were treated within the Perdew--Burke--Ernzerhof (PBE) generalized gradient approximation (GGA) \cite{PBE:ernzerhof1999assessment}. Long-range van der Waals interactions were accounted for using the Grimme DFT-D3 correction with zero damping \cite{grimme2010consistent}. A plane-wave energy cutoff of 550~eV and a $\Gamma$-centered $12\times12\times1$ Monkhorst--Pack $k$-point mesh were employed. A vacuum region of 25~\AA\ was introduced along the $z$ direction to avoid interactions between periodic images. Structural optimizations were carried out until the total-energy convergence reached $10^{-7}$~eV and the residual force on each atom was below 0.01~eV/\AA.
The Heyd--Scuseria--Ernzerhof (HSE06) hybrid functional was used to obtain more accurate electronic band structures and optical properties \cite{HSE:heyd2003hybrid}. Spin--orbit coupling (SOC) effects were also examined and found to have a negligible impact on the calculated band structures and band gaps. Dipole corrections were applied along the out-of-plane direction for the evaluation of electrostatic potentials and work functions, while interfacial charge transfer was quantified using Bader charge analysis. The dynamical stability was examined from phonon-dispersion calculations using density functional perturbation theory (DFPT) on a $3\times3\times1$ supercell with an energy-convergence threshold of $10^{-9}$~eV \cite{DFPT:chaput2011phonon}. The phonon spectra were processed using the PHONOPY package \cite{PHONOPY:togo2023implementation,phonopy-phono3py-JPSJ}. Short-time thermal stability was assessed by ab initio molecular dynamics (AIMD) simulations at 300~K for 5~ps with a time step of 1~fs using the Nos\'e thermostat \cite{MD1:nose1984molecular,MD2:hoover1985canonical}. Mechanical stability was evaluated from the in-plane elastic constants calculated using the stress--strain method. The VASP output files were processed using VASPKIT \cite{vaspkit:wang2021vaspkit}, and the atomic structures and charge-density distributions were visualized using VESTA \cite{VESTA:momma2008vesta}.

\section{Results and discussion}
\subsection{Structure and stability}

Before constructing the heterostructures, the structural properties of the isolated BP, BAs, and Janus WSTe monolayers were optimized. As detailed in the supplementary information, the calculated lattice constants are 3.21, 3.39, and 3.34~\AA, respectively, corresponding to lattice mismatches of about 3.9\% for BP/WSTe and 1.5\% for BAs/WSTe.
For each BX/WSTe ($X$ = P, As) heterostructure, six stacking configurations were considered for both possible orientations of the Janus WSTe layer. Among all configurations, Stacking-4 was found to be the most stable for BP/SWTe, BP/TeWS, BAs/SWTe, and BAs/TeWS, with binding energies of $-0.892$, $-0.930$, $-0.963$, and $-0.998$ eV, respectively. The corresponding interlayer distances are 3.20, 3.41, 3.22, and 3.44~\AA.
The dynamical, short-time thermal, and mechanical stability of the optimized heterostructures was also verified by phonon, AIMD, and elastic-constant calculations. The complete comparison of the stacking configurations and the detailed stability analysis are provided in the Supplementary Information.

The Stacking-4 structures are therefore used in the following electronic, interfacial, optical, and photocatalytic analyses.

\subsection{Electronic properties}

\begin{figure*}[t]
    \centering
    \missinggraphic[width=0.82\textwidth]{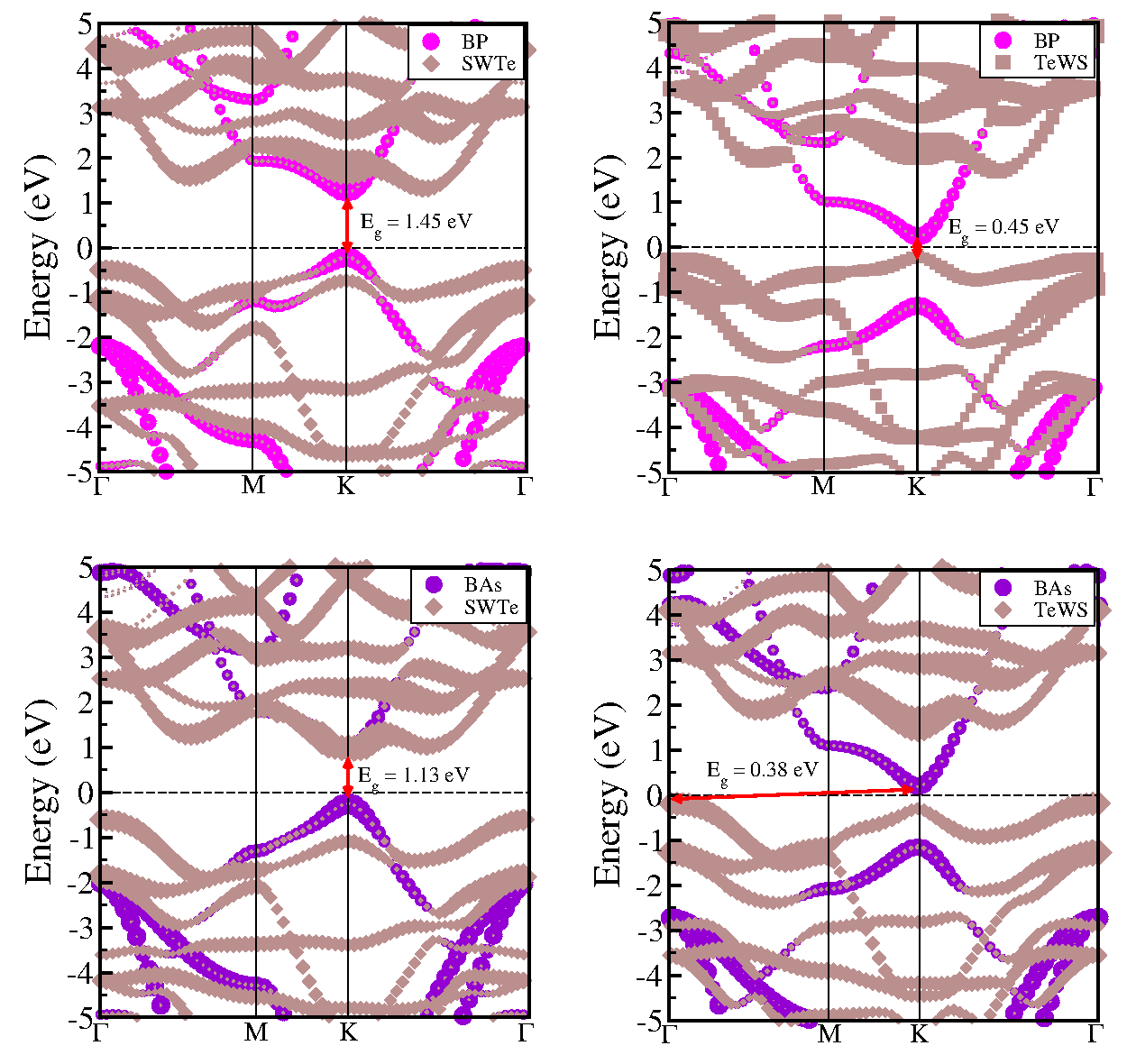}
    \caption{Projected band structures of the BP/SWTe, BP/TeWS, BAs/SWTe, and BAs/TeWS heterostructures. The colored symbols indicate the contributions of the BX and WSTe layers to the electronic states. The energy zero is set at the Fermi level, and the corresponding band-gap values are indicated in each panel.}
    \label{fig:bands}
\end{figure*}

The electronic structures of the isolated BP, BAs, and WSTe monolayers were first calculated using the HSE06 hybrid functional. BP and BAs exhibit direct band gaps of 1.36 and 1.18~eV, respectively, whereas WSTe shows an indirect band gap of 1.59~eV. These values are in good agreement with previously reported theoretical results, as summarized in Figure S1 and Table~S1.
The HSE06 projected band structures of the four investigated heterostructures are shown in Fig.~\ref{fig:bands}. BP/SWTe, BP/TeWS, and BAs/SWTe exhibit direct band gaps of 1.45, 0.45, and 1.13~eV, respectively, whereas BAs/TeWS displays an indirect band gap of 0.38~eV.

The projected band structures further reveal two different band-alignment types. BP/SWTe exhibits a type-I (straddling-gap) alignment, in which both the conduction-band minimum (CBM) and valence-band maximum (VBM) are mainly contributed by the BP layer. Such an alignment favors the confinement of photogenerated electrons and holes within the same layer and is therefore more suitable for optoelectronic and light-emitting applications. This heterostructure will not be considered in the following.
In contrast, BP/TeWS, BAs/SWTe, and BAs/TeWS exhibit type-II (staggered-gap) alignments, where the CBM and VBM are mainly localized on different constituent layers. This spatial separation of the band-edge states is favorable for reducing electron--hole recombination and is therefore more advantageous for photocatalytic applications.
Since the present work focuses on photocatalytic water splitting, the subsequent photocatalytic mechanism and redox analyses focus on the type-II heterostructures BP/TeWS, BAs/SWTe, and BAs/TeWS.

\begin{table*}[htbp]
\centering
\caption{Electronic and interfacial charge-transfer characteristics of the BX/WSTe heterostructures.
$E_g^{\mathrm{HSE}}$ denotes the HSE06 band gap.
$\phi_{\mathrm{high}}$ and $\phi_{\mathrm{low}}$ are the higher and lower surface work functions, respectively, with $\Delta\phi=\phi_{\mathrm{high}}-\phi_{\mathrm{low}}$.
$\Delta Q_{\mathrm{BX}}$ and $\Delta Q_{\mathrm{WSTe}}$ are the Bader charge variations of the two constituent layers; positive (negative) values indicate electron gain (loss).
The interfacial electron-transfer direction and the corresponding junction type are also indicated.}
\label{tab:electronic_charge}

\scriptsize\setlength{\tabcolsep}{3pt}\begin{tabular}{lcccccccc}
\toprule
System &
$E_g^{HSE}$ &
$\phi_{\mathrm{high}}$ &
$\phi_{\mathrm{low}}$ &
$\Delta\phi$ &
$\Delta Q_{\mathrm{BX}}$ &
$\Delta Q_{\mathrm{WSTe}}$ &
Interfacial electron transfer &
Junction type \\
&
(eV) &
(eV) &
(eV) &
(eV) &
($e$) &
($e$) &
& \\
\midrule

BP/SWTe
& 1.45
& 4.83
& 3.64
& 1.19
& $-0.020$
& $+0.020$
& BP $\rightarrow$ WSTe
& Type-I \\

BP/TeWS
& 0.45
& 5.63
& 4.43
& 1.20
& $+0.014$
& $-0.014$
& WSTe $\rightarrow$ BP
& Type-II \\

BAs/SWTe
& 1.13
& 4.68
& 3.55
& 1.13
& $-0.027$
& $+0.027$
& BAs $\rightarrow$ WSTe
& Type-II \\

BAs/TeWS
& 0.38
& 5.53
& 4.30
& 1.23
& $+0.013$
& $-0.013$
& WSTe $\rightarrow$ BAs
& Type-II \\

\bottomrule
\end{tabular}
\end{table*}

\subsection{Interfacial charge transfer}
\begin{figure*}
\centering
\missinggraphic[width=0.9\textwidth]{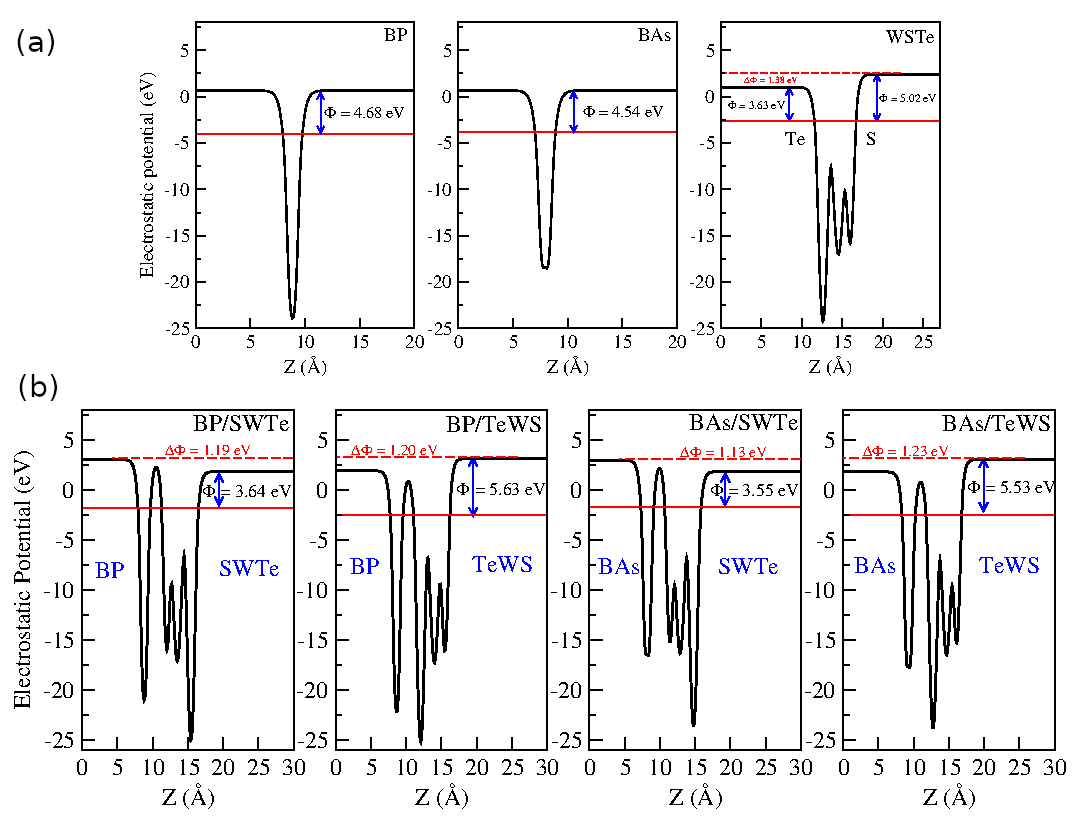}
\caption{Planar-averaged electrostatic potential profiles, work functions, and vacuum level differences of (a) isolated BP, BAs, and WSTe monolayers and (b) BP/SWTe, BP/TeWS, BAs/SWTe, and BAs/TeWS heterostructures.}
\label{fig:02}
\end{figure*}

To understand the origin and direction of the interfacial charge transfer, we first examined the work functions of the isolated constituent monolayers. Upon contact, electrons tend to flow from the material with the lower work function to that with the higher work function until Fermi-level equilibrium is established.
The isolated BP and BAs monolayers exhibit work functions of 4.68 and 4.54~eV, respectively, whereas the Janus WSTe monolayer shows markedly different work functions on its two surfaces, approximately 3.63~eV on the Te side and 5.02~eV on the S side (figure \ref{fig:02}).
Consequently, for the S-terminated interfaces, electrons are expected to transfer from BX toward WSTe, whereas for the Te-terminated interfaces the transfer direction is reversed, from WSTe toward BX.
Such electron redistribution is expected to leave opposite charges on the two constituent layers and to generate a built-in electric field across the interface. For the S-terminated systems, the interfacial electric field is therefore expected to point from BX toward WSTe, whereas for the Te-terminated systems it should point from WSTe toward BX. 

The predicted interfacial charge-transfer directions were further examined and quantified using Bader charge analysis. As listed in Table ~\ref{tab:electronic_charge}, BP and BAs lose 0.020$e$ and 0.027$e$ in BP/SWTe and BAs/SWTe, respectively, with the transferred charge accumulating on WSTe. For BP/TeWS and BAs/TeWS, WSTe instead loses 0.014$e$ and 0.013$e$, respectively, and the charge is transferred to the BX layer.

Notably, the transferred charge remains relatively small, consistent with the predominantly van der Waals nature of the interlayer interaction, while still indicating sufficient interfacial electronic coupling to establish the built-in internal electric field and promote carrier migration.
In addition to this IEF, the asymmetric nature of the complete heterostructures gives rise to different vacuum
levels at their two outer surfaces (fig. \ref{fig:02}.) This out-of-plane electrostatic asymmetry is characterized by the surface potential difference, $\Delta\phi$. The calculated values are 1.19, 1.20, 1.13, and 1.23~eV for BP/SWTe, BP/TeWS, BAs/SWTe, and BAs/TeWS, respectively (table \ref{tab:electronic_charge}). It is important to distinguish the surface potential difference, $\Delta\phi$, from the interfacial built-in electric field (IEF). Here, $\Delta\phi$ corresponds to the difference between the vacuum levels on the two outer sides of the heterostructure, whereas the IEF originates from charge redistribution between the BX and WSTe layers at the interface. The calculated $\Delta\phi$ is later used in the STH efficiency correction.

\begin{figure*}
\centering
\missinggraphic[width=0.8\textwidth]{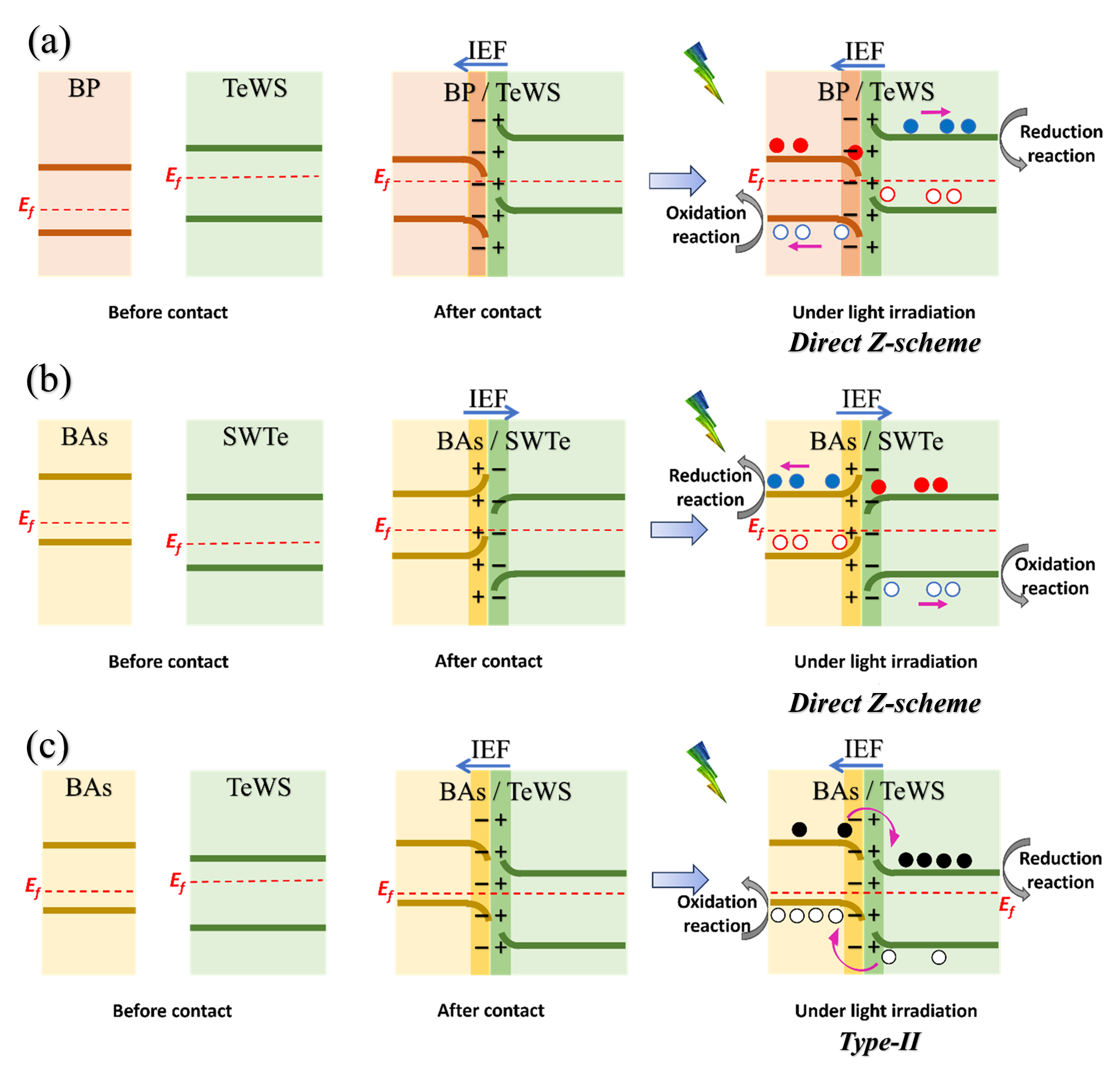}
\caption{Schematic illustrations of the interfacial charge-transfer mechanisms in the BX/WSTe heterostructures: (a) BP/TeWS direct Z-scheme, (b) BAs/SWTe direct Z-scheme, and (c) BAs/TeWS conventional type-II mechanism.}
\label{fig:04}
\end{figure*}

To further analyze the carrier-transfer mechanism inferred from the work function analysis and confirmed by the Bader charge redistribution, schematic illustrations of the interfacial carrier migration pathways are presented in Fig.~\ref{fig:04}.

As shown in Fig.~\ref{fig:04}(a), the electronic structure of the BP/TeWS heterostructure is consistent with a direct Z-scheme charge-transfer mechanism. Before contact, the two constituent monolayers possess different Fermi levels. After contact, Fermi-level equilibration leads to interfacial charge redistribution and the formation of a built-in electric field directed from TeWS toward BP. The resulting band bending originates from charge accumulation at the interface. Under illumination, this interfacial electrostatic environment is expected to favor the recombination of the less reactive photogenerated carriers at the interface, leaving the electrons with stronger reduction capability in the TeWS layer and the holes with stronger oxidation capability in the BP layer. Such a carrier-transfer pathway is consistent with a direct Z-scheme mechanism and preserves the high redox ability required for photocatalytic reactions.

The electronic and interfacial characteristics of BAs/SWTe are also consistent with a direct Z-scheme charge-transfer mechanism, as illustrated in Fig.~\ref{fig:04}(b). In this case, however, the IEF and interfacial charge redistribution favor the recombination of the photogenerated electrons in the SWTe conduction band with holes in the BAs valence band. Consequently, the electrons with strong reduction capability remain in BAs, whereas the strongly oxidative holes remain in SWTe. This charge-transfer pathway also enables spatial separation of the active photogenerated carriers while retaining their strong redox potentials.

By contrast, the BAs/TeWS heterostructure follows a conventional type-II charge-transfer mechanism, as shown in Fig.~\ref{fig:04}(c). Under light irradiation, photogenerated electrons migrate from BAs to TeWS, while holes transfer in the opposite direction. This process promotes spatial charge separation, but the remaining electrons and holes occupy lower-energy reduction and oxidation states, respectively, resulting in a partial loss of redox power compared with the direct Z-scheme systems.

\subsection{Optical absorptance properties}

Efficient utilization of solar radiation is an essential requirement for photocatalytic water splitting. To evaluate the optical response of the BP/SWTe, BP/TeWS, BAs/SWTe, and BAs/TeWS heterostructures, we calculated the frequency-dependent complex dielectric function within the independent-particle approximation. Within this framework, the imaginary part of the dielectric function originates from interband electronic transitions, whereas the real part is obtained through the Kramers--Kronig relation~\cite{gajdos2006linear}. The complex in-plane dielectric function can be written as

\begin{equation}
\varepsilon_\parallel(\omega)
=
\varepsilon_{1\parallel}(\omega)
+
i\varepsilon_{2\parallel}(\omega),
\end{equation}

where $\varepsilon_{1\parallel}(\omega)$ and $\varepsilon_{2\parallel}(\omega)$ are the real and imaginary parts, respectively.
\begin{figure*}
\centering
\missinggraphic[width=1\textwidth]{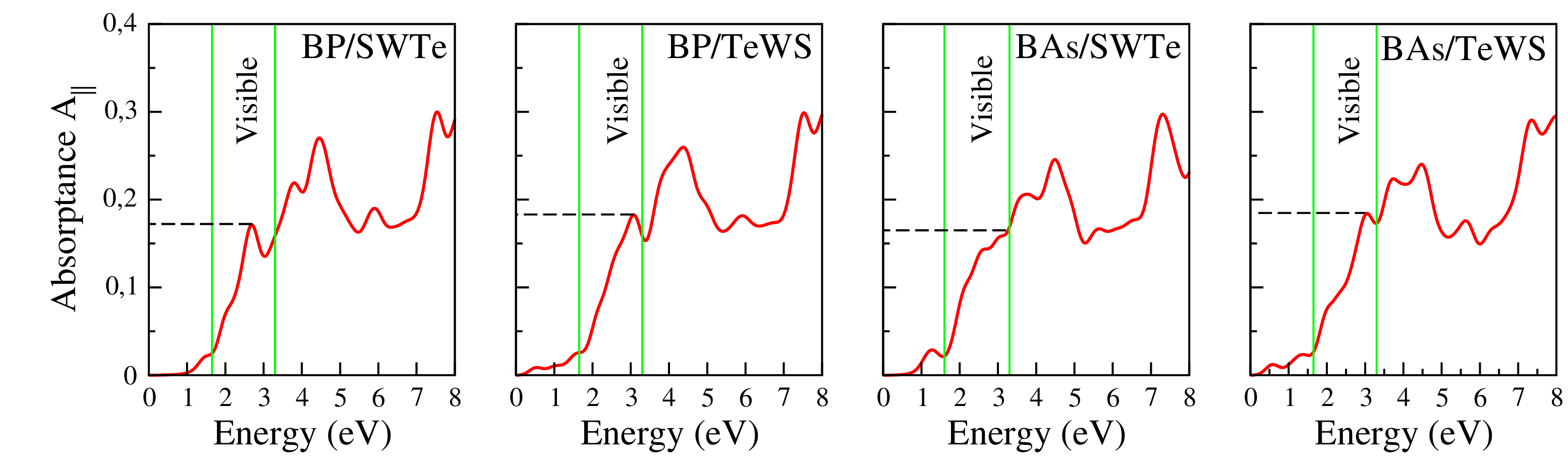}
\caption{In-plane absorptance spectra, $A_{\parallel}$, of the BP/SWTe, BP/TeWS, BAs/SWTe, and BAs/TeWS heterostructures. The horizontal dashed lines indicate the maximum absorptance within the visible-light region.}
\label{fig:05}
\end{figure*}
For an atomically thin material described in a supercell approach, the dielectric response obtained directly from the DFT calculation can be considered as an effective dielectric function of a fictitious homogeneous material filling the supercell. It does not directly represent the fraction of incident optical power absorbed by a single layer. As a consequence, the dielectric function obtained from a periodically repeated two-dimensional system depends on the vacuum spacing used in the simulation cell and is therefore not an intrinsic property of the 2D material. A more suitable description of the optical response of an atomically thin material is therefore obtained by treating it as an optical sheet characterized by a two-dimensional sheet conductivity~\cite{li2018twodimensional,majerus2023anisotropy}.

The dielectric response of the periodic supercell was converted into the two-dimensional sheet conductivity according to~\cite{majerus2023anisotropy}

\begin{equation}
\sigma_{2D}(\omega)
=
-i\varepsilon_0\omega L_z
\left[
\varepsilon_\parallel(\omega)-1
\right],
\end{equation}

where $\varepsilon_0$ is the vacuum permittivity and $L_z$ is the supercell length normal to the two-dimensional plane.

For convenience, a dimensionless optical-response function is introduced as

\begin{equation}
g(\omega)
=
Z_0\sigma_{2D}(\omega),
\end{equation}

where $Z_0$ is the impedance of free space. For a free-standing two-dimensional sheet under normal incidence, the absorptance can be expressed as~\cite{li2018twodimensional,lobet2016perfect}

\begin{equation}
A_\parallel(\omega)
=
\frac{
4\,\mathrm{Re}[g(\omega)]
}{
|2+g(\omega)|^2
}.
\end{equation}

The resulting $A_{\parallel}$ spectra are displayed in Fig.~\ref{fig:05}, while the horizontal dashed lines indicate the maximum absorptance reached within the visible-light region. $A_{\parallel}$ directly represents the fraction of incident optical power absorbed by a single heterostructure layer. The calculated single-layer absorptance remains well below unity, as expected for an atomically thin material. This is relevant to the STH calculation because the conventional model effectively assumes complete harvesting of photons above the band gap. In the following section, the calculated $A_{\parallel}$ is therefore used to weight the AM1.5G solar spectrum in the light-harvesting term.

\subsection{Photocatalytic mechanism}

\subsubsection{Band-edge alignment and redox capability}

Having established the interfacial charge-transfer characteristics of the BX/WSTe heterostructures, we next examine whether the photogenerated carriers that accumulate at the active band edges possess sufficient thermodynamic driving force to promote the hydrogen evolution reaction (HER) and oxygen evolution reaction (OER). Accordingly, only the band-edge states identified in the previous section as the accumulation sites of photogenerated electrons and holes were considered. These active electron and hole states were aligned with the vacuum level and compared with the pH-dependent H$^+$/H$_2$ and O$_2$/H$_2$O redox potentials. This representation therefore focuses directly on the charge carriers available for the photocatalytic redox reactions and allows the thermodynamically accessible pH range for overall water splitting to be determined.

\begin{figure*}
\centering
\missinggraphic[width=0.8\textwidth]{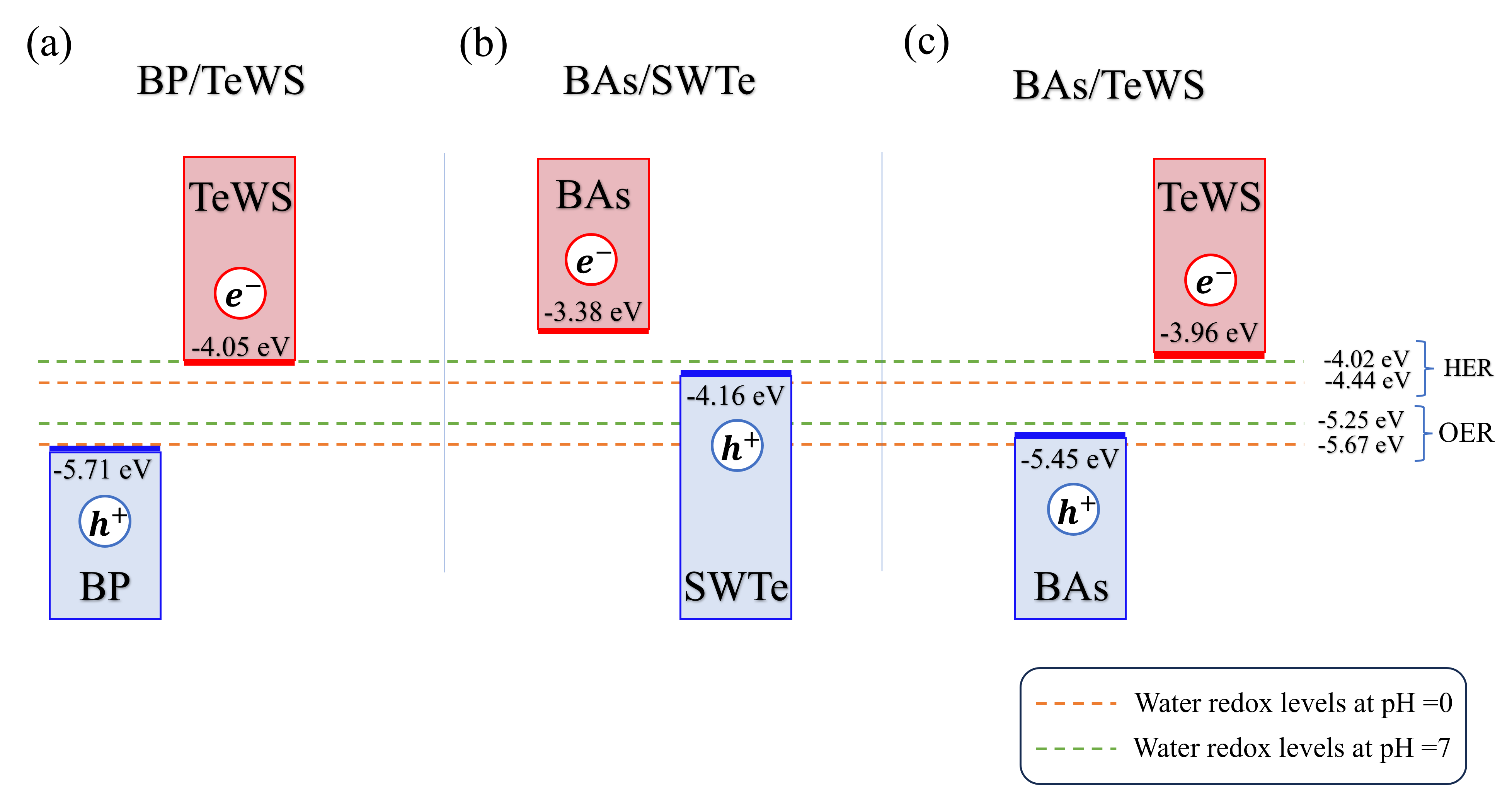}
\caption{Active band-edge positions of BP/TeWS, BAs/SWTe, and BAs/TeWS relative to the vacuum level. The pH-dependent thermodynamic water-splitting ranges are determined by comparing these active band edges with the H$^+$/H$_2$ and O$_2$/H$_2$O redox potentials.}
\label{fig:06}
\end{figure*}
The reduction and oxidation potentials of water vary with pH according to \cite{10.1021/acsomega.0c02804}

\begin{equation}
E_{\mathrm{H^+/H_2}}
=
-4.44 + 0.059\,\mathrm{pH}
\quad \mathrm{eV},
\end{equation}

and

\begin{equation}
E_{\mathrm{O_2/H_2O}}
=
-5.67 + 0.059\,\mathrm{pH}
\quad \mathrm{eV}.
\end{equation}

For overall water splitting, the active electron state must lie at a higher energy than the H$^+$/H$_2$ reduction potential, whereas the active hole state must lie at a lower energy than the O$_2$/H$_2$O oxidation potential. Since both redox potentials shift with pH, these conditions define a specific operating pH range for each heterostructure.

As illustrated in Fig. 5, the active electron and hole states of BP/TeWS are located at $-4.05$ and $-5.71$ eV, respectively. Comparison with the pH-dependent water redox potentials shows that the HER condition is satisfied for $\mathrm{pH}<6.6$, while the OER condition remains satisfied throughout this interval. Consequently, BP/TeWS can thermodynamically support overall water splitting approximately within the range
\[
0 < \mathrm{pH} < 6.6.
\]
For BAs/TeWS, the active electron and hole states are located at $-3.96$ and $-5.45$ eV, respectively. The HER condition requires $\mathrm{pH}<8.1$, whereas the OER becomes thermodynamically accessible for $\mathrm{pH}>3.7$. Therefore, BAs/TeWS supports overall water splitting within the range
\[
3.7 < \mathrm{pH} < 8.1.
\] 
In contrast, BAs/SWTe exhibits an asymmetric redox capability. Although its active electron state at $-3.38$ eV provides sufficient driving force for the HER throughout the conventional aqueous pH range, its active hole state at $-4.16$ eV does not provide sufficient oxidation power for the OER. Consequently, BAs/SWTe cannot independently promote overall water splitting within the conventional aqueous pH range.

\subsubsection{Solar-to-hydrogen conversion efficiency}
To quantitatively evaluate the solar-to-hydrogen (STH) conversion capability of the photocatalytically active heterostructures, we employed the theoretical framework proposed by Fu et al.~\cite{fu2018intrinsic}. This approach provides a theoretical estimate of the STH efficiency from the AM1.5G solar spectrum together with the electronic and redox properties of the photocatalyst.

The STH efficiency is expressed as
\begin{equation}
\eta_{\mathrm{STH}}
=
\eta_{\mathrm{abs}}
\times
\eta_{\mathrm{cu}},
\end{equation}

where $\eta_{\mathrm{abs}}$ and $\eta_{\mathrm{cu}}$ represent the light-absorption and carrier-utilization efficiencies, respectively.

The light-absorption efficiency is given by

\begin{equation}
\eta_{\mathrm{abs}}
=
\frac{
\displaystyle\int_{E_g}^{\infty}
A_{\parallel}(\hbar\omega)
P(\hbar\omega)\,d(\hbar\omega)
}{
\displaystyle\int_{0}^{\infty}
P(\hbar\omega)\,d(\hbar\omega)
},
\end{equation}

where $P(\hbar\omega)$ denotes the AM1.5G solar spectral irradiance taken from the ASTM G173 reference spectrum~\cite{ASTM_G173}, $E_g$ is the band gap of the photocatalyst, and $A_{\parallel}(\hbar\omega)$ is the in-plane absorptance.

The carrier-utilization efficiency is given by

\begin{equation}
\eta_{\mathrm{cu}}
=
\frac{
\Delta G
\displaystyle\int_{E}^{\infty}
A_{\parallel}(\hbar\omega)
\frac{P(\hbar\omega)}{\hbar\omega}
\,d(\hbar\omega)
}{
\displaystyle\int_{E_g}^{\infty}
A_{\parallel}(\hbar\omega)
P(\hbar\omega)\,d(\hbar\omega)
},
\end{equation}

where $\Delta G=1.23$~eV is the thermodynamic energy required for overall water splitting, i.e. the difference between the redox potential of the HER and the OER. The minimum photon energy, $E$, that can effectively contribute to the reaction additionally depends on the available driving forces for the hydrogen and oxygen evolution reactions, $\chi_{H_2}$ and $\chi_{O_2}$, respectively. Here, $\chi_{H_2}$ represents the energy difference between the active conduction-band edge and the H$^+$/H$_2$ reduction potential, while $\chi_{O_2}$ represents the energy difference between the active valence-band edge and the O$_2$/H$_2$O oxidation potential.

\begin{equation}
E = \begin{cases}
E_g & (\chi_{H_2} \ge 0.2, \chi_{O_2} \ge 0.6) \\
E_g + 0.2 - \chi_{H_2}
& (\chi_{H_2} < 0.2, \chi_{O_2} \ge 0.6) \\
E_g + 0.6 - \chi_{O_2}
& (\chi_{H_2} \ge 0.2, \chi_{O_2} < 0.6) \\
E_g + 0.8 - \chi_{H_2} - \chi_{O_2}
& (\chi_{H_2} < 0.2, \chi_{O_2} < 0.6).
\end{cases}
\end{equation}

Here, 0.2 and 0.6~eV correspond to the energetic requirements adopted for the HER and OER, respectively.

For photocatalysts possessing an intrinsic out-of-plane potential difference, Fu et al. further introduced a correction accounting for the vacuum-level difference, $\Delta\phi$, between the two opposite surfaces:

\begin{equation}
\eta'_{\mathrm{STH;A}}
=
\eta_{\mathrm{STH;A}}
\frac{
\displaystyle\int_{0}^{\infty}
P(\hbar\omega)\,d(\hbar\omega)
}{
\displaystyle\int_{0}^{\infty}
P(\hbar\omega)\,d(\hbar\omega)
+
\Delta\phi
\displaystyle\int_{E_g}^{\infty}
A_{\parallel}(\hbar\omega)
\frac{P(\hbar\omega)}{\hbar\omega}
\,d(\hbar\omega)
}.
\end{equation}

In the conventional Fu formulation, complete light harvesting above the band gap is assumed, corresponding to $A_{\parallel}(\hbar\omega)=1$ for $\hbar\omega \geq E_g$~\cite{fu2018intrinsic}. We then define

\begin{equation}
\eta'_{\mathrm{STH;Fu}}
=
\left.
\eta'_{\mathrm{STH;A}}
\right|_{A_{\parallel}(\hbar\omega)=1,\;\hbar\omega\geq E_g}.
\end{equation}

Hereafter, $\eta'_{\mathrm{STH;Fu}}$ denotes the STH
efficiency obtained under the ideal light-harvesting
assumption, whereas $\eta'_{\mathrm{STH;A}}$
represents the corresponding efficiency calculated
using the actual energy-dependent absorptance. Both quantities include the correction for the out-of-plane potential difference, $\Delta\phi$.

In contrast to the idealized full-absorption approximation, the photon-flux contribution entering the surface-potential correction is weighted by the energy-dependent absorptance. This extension accounts for the fact that only absorbed photons can generate photocarriers, thereby providing a consistent treatment of light harvesting throughout the efficiency model.

It is important to note that the STH efficiencies obtained from such theoretical models represent idealized limits and are often considerably higher than experimentally achieved values. Experimental overall water-splitting systems still suffer from several additional loss mechanisms, including incomplete light harvesting, carrier recombination, sluggish surface reaction kinetics, and transport losses.
Fu et al. reported theoretical STH efficiencies, $\eta'_{\mathrm{STH;Fu}}$, reaching several tens of percent for various two-dimensional photocatalysts within the idealized light-harvesting framework~\cite{fu2018intrinsic}. More recently, Liu et al. reported a theoretical STH efficiency of 27.54\% for the SnSSe/$\beta$-C$_{20}$ heterostructure~\cite{liu2025two}, while Cheng et al. predicted corrected STH efficiencies, $\eta'_{\mathrm{STH;Fu}}$, of 25.42\% and 25.71\% for two Sc$_2$CBr$_2$/MoTe$_2$ stacking configurations under 4\% tensile strain, respectively~\cite{cheng2025tunable}.

By comparison, a highly optimized InGaN-based photocatalytic system experimentally achieved a much lower STH efficiency of 9.2\% under concentrated solar illumination~\cite{zhou2023sth}, while a more recent halide-perovskite-based overall water-splitting system reached 2.07\%~\cite{fu2024perovskite}. These experimental values should not be regarded as direct benchmarks for the present two-dimensional heterostructures because the materials, architectures, and operating conditions are different; nevertheless, they illustrate the large difference that can exist between ideal theoretical estimates and experimentally realized photocatalytic efficiencies.

\begin{figure*}
\centering
\missinggraphic[width=0.8\textwidth]{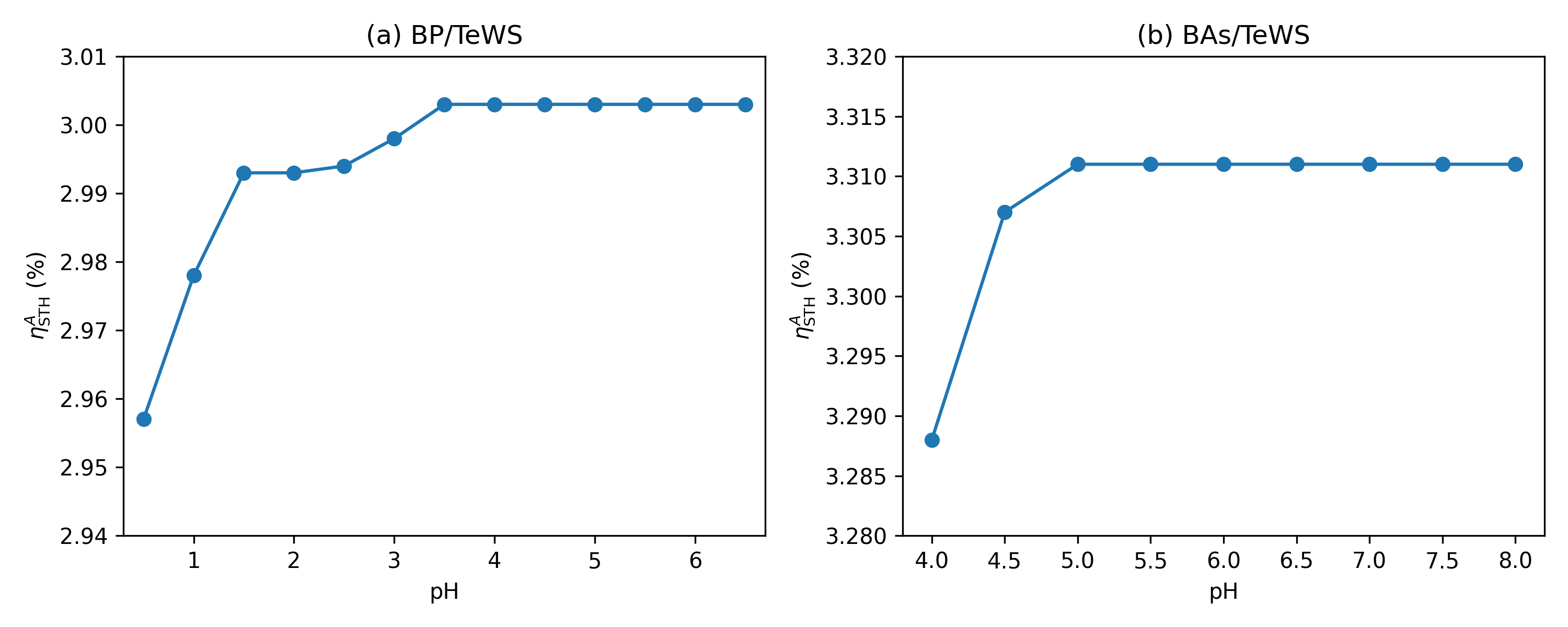}
\caption{Variation of the absorptance-weighted solar-to-hydrogen efficiency ($\eta'_{\mathrm{STH;A}}$) with pH for (a) BP/TeWS and (b) BAs/TeWS heterostructures. The calculated efficiencies are shown within the thermodynamically allowed pH ranges for overall water splitting. Symbols denote calculated values, and solid lines are guides to the eye.}
\label{fig:sth_comparison}
\end{figure*}

Using the calculated actual absorptance of the 2D materials, we have analysed the corrected STH efficiency, $\eta'_{\mathrm{STH;A}}$, as a function of pH. For each pH value, $\chi_{H_2}$ and $\chi_{O_2}$ were recalculated from the corresponding redox potentials, leading to a pH-dependent minimum usable photon energy, $E$. The resulting $\eta'_{\mathrm{STH;A}}$ values are shown in Fig.~\ref{fig:sth_comparison}, while the complete numerical data are provided in the Supplementary Information.
For BP/TeWS, the corrected absorptance-weighted STH efficiency increases slightly from approximately 2.87\% at pH 0.5 to 2.91\% at pH 3.5, remaining nearly constant up to the upper thermodynamic limit of pH $\approx$ 6.6. For BAs/TeWS, the corresponding efficiency increases from approximately 3.18\% at pH 4 to 3.20\% at pH 5, remaining nearly unchanged up to pH $\approx$ 8.1.
Thus, although the operating pH ranges of the two heterostructures differ, the absorptance-weighted STH efficiency varies only weakly within the thermodynamically accessible range of each system. The substantial reduction compared with the ideal Fu formulation highlights the importance of accounting for the actual optical absorptance of atomically thin photocatalysts.

\section*{Conclusions}
In this work, the structural, electronic, interfacial, optical, and photocatalytic properties of BX/WSTe ($X$ = P, As) heterostructures were investigated using first-principles calculations. The results demonstrate that reversing the orientation of the Janus WSTe layer changes the direction of interfacial charge transfer, leading to distinct carrier-transfer mechanisms. Based on the band alignment and interfacial charge redistribution, BP/TeWS and BAs/SWTe exhibit characteristics consistent with direct Z-scheme charge transfer, whereas BAs/TeWS follows a conventional type-II mechanism.

Band-edge analysis indicates that BP/TeWS can thermodynamically support overall water splitting at pH values below approximately 6.6, whereas BAs/TeWS is suitable within $3.7 < \mathrm{pH} < 8.1$. In contrast, BAs/SWTe supports only the HER within the conventional aqueous pH range.
The absorptance-weighted STH efficiency, $\eta'_{\mathrm{STH;A}}$, reaches approximately 2.87--2.91\% for BP/TeWS and 3.17--3.20\% for BAs/TeWS. These values are substantially lower than those predicted under the assumption of complete light harvesting, highlighting the importance of accounting for the actual optical absorptance of atomically thin photocatalysts.

Nevertheless, the present assessment is primarily based on thermodynamic band-edge alignment and does not explicitly account for reaction kinetics, surface adsorption, charge-carrier recombination, or solvent effects. Further investigations incorporating these factors are therefore required to assess the practical photocatalytic performance of the proposed heterostructures.

Overall, this work highlights Janus-based heterostructures as a versatile platform for controlling interfacial charge transfer through the orientation of the polar layer. The proposed methodology, combining first-principles electronic and interfacial properties with absorptance-weighted STH efficiency calculations, can be extended to other atomically thin photocatalysts.

\section*{Declaration of competing interest}
There are no conflicts to declare.

\section*{Acknowledgements}
This research used resources of the "Plateforme Technologique de Calcul Intensif (PTCI)" located at the University of Namur, Belgium, which is supported  by the FNRS-FRFC, the Walloon Region, and the University of Namur (Conventions No. 2.5020.11, GEQ U.G006.15, 1610468, RW/GEQ2016 et U.G011.22). The PTCI is member of the "Consortium des \'{E}quipements de Calcul Intensif  (C\'{E}CI)"

\bibliographystyle{unsrt}
\bibliography{rsc}

@article{BX:islam2022tuning,
  title={Tuning the electronic, phonon, and optical properties of monolayer BX (XP and As) through the strain effect},
  author={Islam, Rasidul and Islam, Sherajul and Mojumder, Rayid Hasan and Khan, Zarif and Molla, Hasan and Islam, ASM Jannatul and Park, Jeongwon},
  journal={Materials Today Communications},
  volume={33},
  pages={104227},
  year={2022},
  publisher={Elsevier}
}

@article{BX:csahin2009monolayer,
  title={Monolayer honeycomb structures of group-IV elements and III-V binary compounds: First-principles calculations},
  author={{\c{S}}ahin, Hasan and Cahangirov, Seymur and Topsakal, Mehmet and Bekaroglu, E and Akturk, Ethem and Senger, R Tugrul and Ciraci, Salim},
  journal={Physical Review B—Condensed Matter and Materials Physics},
  volume={80},
  number={15},
  pages={155453},
  year={2009},
  publisher={APS}
}

@article{BX:khurami2023strain,
  title={Strain engineering on structural, optoelectronic and photocatalytic properties of bp, bas and bsb monolayers},
  author={Khurami, MW and Amin, B and Chen, Yuanping and Yan, Xiaohong and Idrees, M},
  journal={Solid State Communications},
  volume={371},
  pages={115273},
  year={2023},
  publisher={Elsevier}
}

@article{WSTe:patel2020high,
  title={High thermoelectric performance in two-dimensional Janus monolayer material WS-X (X= Se and Te)},
  author={Patel, Abhishek and Singh, Deobrat and Sonvane, Yogesh and Thakor, PB and Ahuja, Rajeev},
  journal={ACS applied materials \& interfaces},
  volume={12},
  number={41},
  pages={46212--46219},
  year={2020},
  publisher={ACS Publications}
}

@article{mechstab:wang2022high,
  title={High-throughput computational screening of two-dimensional semiconductors},
  author={Wang, Vei and Tang, Gang and Liu, Ya-Chao and Wang, Ren-Tao and Mizuseki, Hiroshi and Kawazoe, Yoshiyuki and Nara, Jun and Geng, Wen Tong},
  journal={The Journal of Physical Chemistry Letters},
  volume={13},
  number={50},
  pages={11581--11594},
  year={2022},
  publisher={ACS Publications}
}

@book{stress:born1996dynamical,
  title={Dynamical theory of crystal lattices},
  author={Born, Max and Huang, Kun},
  year={1996},
  publisher={Oxford university press}
}

@article{fu2018intrinsic,
  title={Intrinsic electric fields in two-dimensional materials boost the solar-to-hydrogen efficiency for photocatalytic water splitting},
  author={Fu, Cen-Feng and Sun, Jiuyu and Luo, Qiquan and Li, Xingxing and Hu, Wei and Yang, Jinlong},
  journal={Nano letters},
  volume={18},
  number={10},
  pages={6312--6317},
  year={2018},
  publisher={ACS Publications}
}

@article{gao2019zno,
  title={ZnO/g-GeC van der Waals heterostructure: novel photocatalyst for small molecule splitting},
  author={Gao, Xu and Shen, Yanqing and Ma, Yanyan and Wu, Shengyao and Zhou, Zhongxiang},
  journal={Journal of Materials Chemistry C},
  volume={7},
  number={16},
  pages={4791--4799},
  year={2019},
  publisher={Royal Society of Chemistry}
}

@article{niu2017arsenene,
  title={Arsenene-based heterostructures: highly efficient bifunctional materials for photovoltaics and photocatalytics},
  author={Niu, Xianghong and Li, Yunhai and Zhou, Qionghua and Shu, Huabing and Wang, Jinlan},
  journal={ACS applied materials \& interfaces},
  volume={9},
  number={49},
  pages={42856--42861},
  year={2017},
  publisher={ACS Publications}
}

@article{ran2021significantly,
  title={Significantly raised visible-light photocatalytic H2 evolution on a 2D/2D ReS2/In2ZnS4 van der Waals heterostructure},
  author={Ran, Jingrun and Zhang, Hongping and Qu, Jiangtao and Shan, Jieqiong and Davey, Kenneth and Cairney, Julie M and Jing, Liqiang and Qiao, Shi-Zhang},
  journal={Small},
  volume={17},
  number={32},
  pages={2100296},
  year={2021},
  publisher={Wiley Online Library}
}

@article{li2018type,
  title={Type-II InSe/MoSe 2 (WSe 2) van der Waals heterostructures: vertical strain and electric field effects},
  author={Li, Xueping and Jia, Guangrui and Du, Juan and Song, Xiaohui and Xia, Congxin and Wei, Zhongming and Li, Jingbo},
  journal={Journal of Materials Chemistry C},
  volume={6},
  number={37},
  pages={10010--10019},
  year={2018},
  publisher={Royal Society of Chemistry}
}

@article{zhou2020type,
  title={Type-II/type-II band alignment to boost spatial charge separation: a case study of gC 3 N 4 quantum dots/a-TiO 2/r-TiO 2 for highly efficient photocatalytic hydrogen and oxygen evolution},
  author={Zhou, Bing-Xin and Ding, Shuang-Shuang and Wang, Yan and Wang, Xiao-Rui and Huang, Wei-Qing and Li, Kai and Huang, Gui-Fang},
  journal={Nanoscale},
  volume={12},
  number={10},
  pages={6037--6046},
  year={2020},
  publisher={Royal Society of Chemistry}
}

@article{mao2021hierarchically,
  title={Hierarchically 1D CdS decorated on 2D perovskite-type La2Ti2O7 nanosheet hybrids with enhanced photocatalytic performance},
  author={Mao, Liang and Cai, Xiao-Yan and Zhu, Ming-Shan},
  journal={Rare Metals},
  volume={40},
  number={5},
  pages={1067--1076},
  year={2021},
  publisher={Springer}
}

@article{huang2017switching,
  title={Switching charge transfer of C3N4/W18O49 from type-II to Z-scheme by interfacial band bending for highly efficient photocatalytic hydrogen evolution},
  author={Huang, Zhen-Feng and Song, Jiajia and Wang, Xin and Pan, Lun and Li, Ke and Zhang, Xiangwen and Wang, Li and Zou, Ji-Jun},
  journal={Nano Energy},
  volume={40},
  pages={308--316},
  year={2017},
  publisher={Elsevier}
}

@article{low2017review,
  title={A review of direct Z-scheme photocatalysts},
  author={Low, Jingxiang and Jiang, Chuanjia and Cheng, Bei and Wageh, Swelm and Al-Ghamdi, Ahmed A and Yu, Jiaguo},
  journal={Small methods},
  volume={1},
  number={5},
  pages={1700080},
  year={2017},
  publisher={Wiley Online Library}
}

@article{xu2018direct,
  title={Direct Z-scheme photocatalysts: Principles, synthesis, and applications},
  author={Xu, Quanlong and Zhang, Liuyang and Yu, Jiaguo and Wageh, Swelm and Al-Ghamdi, Ahmed A and Jaroniec, Mietek},
  journal={Materials Today},
  volume={21},
  number={10},
  pages={1042--1063},
  year={2018},
  publisher={Elsevier}
}

@article{meng2017direct,
  title={Direct Z-scheme TiO2/CdS hierarchical photocatalyst for enhanced photocatalytic H2-production activity},
  author={Meng, Aiyun and Zhu, Bicheng and Zhong, Bo and Zhang, Liuyang and Cheng, Bei},
  journal={Applied Surface Science},
  volume={422},
  pages={518--527},
  year={2017},
  publisher={Elsevier}
}

@article{jo2015influence,
  title={Influence of TiO2 morphology on the photocatalytic efficiency of direct Z-scheme g-C3N4/TiO2 photocatalysts for isoniazid degradation},
  author={Jo, Wan-Kuen and Natarajan, Thillai Sivakumar},
  journal={Chemical Engineering Journal},
  volume={281},
  pages={549--565},
  year={2015},
  publisher={Elsevier}
}

@article{wang2018van,
  title={Van der Waals heterostructures comprised of ultrathin polymer nanosheets for efficient Z-scheme overall water splitting},
  author={Wang, Lei and Zheng, Xusheng and Chen, Liang and Xiong, Yujie and Xu, Hangxun},
  journal={Angewandte Chemie International Edition},
  volume={57},
  number={13},
  pages={3454--3458},
  year={2018},
  publisher={Wiley Online Library}
}

@article{hu2019z,
  title={Z-Scheme 2D/2D heterojunction of black phosphorus/monolayer Bi2WO6 nanosheets with enhanced photocatalytic activities},
  author={Hu, Jundie and Chen, Dongyun and Mo, Zhao and Li, Najun and Xu, Qingfeng and Li, Hua and He, Jinghui and Xu, Hui and Lu, Jianmei},
  journal={Angewandte Chemie International Edition},
  volume={58},
  number={7},
  pages={2073--2077},
  year={2019},
  publisher={Wiley Online Library}
}

@article{niu2020rational,
  title={Rational design and characterization of direct Z-scheme photocatalyst for overall water splitting from excited state dynamics simulations},
  author={Niu, Xianghong and Bai, Xiaowan and Zhou, Zhaobo and Wang, Jinlan},
  journal={ACS catalysis},
  volume={10},
  number={3},
  pages={1976--1983},
  year={2020},
  publisher={ACS Publications}
}

@article{wang2021constructing,
  title={Constructing a new 2D Janus black phosphorus/SMoSe heterostructure for spontaneous wide-spectral-responsive photocatalytic overall water splitting},
  author={Wang, Guoqing and Chen, Chen and Teketel, Birkneh Sirak and Xu, Bo and Lin, Bin},
  journal={International Journal of Hydrogen Energy},
  volume={46},
  number={79},
  pages={39183--39194},
  year={2021},
  publisher={Elsevier}
}

@article{tang2023engineering,
  title={Engineering two-dimensional SnC/HfSSe heterojunction as a direct Z-scheme photocatalyst for water splitting hydrogen evolution},
  author={Tang, Wenyi and Wang, Guangzhao and Fu, Can and Wang, Biao and Yuan, Hongkuan and Chen, Hong},
  journal={Applied Surface Science},
  volume={626},
  pages={157247},
  year={2023},
  publisher={Elsevier}
}

@article{alrebdi2022situ,
  title={In-situ formation of Are-MXY (M= Mo, W;(X$\ne$ Y)= S, Se, Te) van der Waals heterostructure},
  author={Alrebdi, Tahani A and Idrees, M and Alkallas, FH and Amin, B},
  journal={Journal of Solid State Chemistry},
  volume={313},
  pages={123284},
  year={2022},
  publisher={Elsevier}
}

@article{zhang2017janus,
  title={Janus monolayer transition-metal dichalcogenides},
  author={Zhang, Jing and Jia, Shuai and Kholmanov, Iskandar and Dong, Liang and Er, Dequan and Chen, Weibing and Guo, Hua and Jin, Zehua and Shenoy, Vivek B and Shi, Li and others},
  journal={ACS nano},
  volume={11},
  number={8},
  pages={8192--8198},
  year={2017},
  publisher={ACS Publications}
}

@article{hernandez2022janus,
  title={Janus monolayers of transition metal dichalcogenides: A DFT study},
  author={Hern{\'a}ndez-V{\'a}zquez, Miguel {\'A}ngel and de Luna Bugallo, Andr{\'e}s and Olgu{\'\i}n, Daniel},
  journal={physica status solidi (b)},
  volume={259},
  number={1},
  pages={2100248},
  year={2022},
  publisher={Wiley Online Library}
}

@article{riis2019classifying,
  title={Classifying the electronic and optical properties of Janus monolayers},
  author={Riis-Jensen, Anders C and Deilmann, Thorsten and Olsen, Thomas and Thygesen, Kristian S},
  journal={ACS nano},
  volume={13},
  number={11},
  pages={13354--13364},
  year={2019},
  publisher={ACS Publications}
}

@article{idrees2020van,
  title={Van der Waals heterostructures of SiC and Janus MSSe (M= Mo, W) monolayers: a first principles study},
  author={Idrees, M and Fawad, M and Bilal, M and Saeed, Y and Nguyen, C and Amin, Bin},
  journal={RSC advances},
  volume={10},
  number={43},
  pages={25801--25807},
  year={2020},
  publisher={Royal Society of Chemistry}
}

@article{wang2022two,
  title={Two-dimensional CdO/PtSSe heterojunctions used for Z-scheme photocatalytic water-splitting},
  author={Wang, Guangzhao and Tang, Wenyi and Xu, Chunping and He, Jing and Zeng, Qiurong and Xie, Wenjie and Gao, Peng and Chang, Junli},
  journal={Applied Surface Science},
  volume={599},
  pages={153960},
  year={2022},
  publisher={Elsevier}
}

@article{idrees2019optoelectronic,
  title={Optoelectronic and solar cell applications of Janus monolayers and their van der Waals heterostructures},
  author={Idrees, M and Din, HU and Ali, R and Rehman, G and Hussain, T and Nguyen, CV and Ahmad, Iftikhar and Amin, B},
  journal={Physical Chemistry Chemical Physics},
  volume={21},
  number={34},
  pages={18612--18621},
  year={2019},
  publisher={Royal Society of Chemistry}
}

@article{vasp1:kresse1993ab,
  title={Ab initio molecular dynamics for open-shell transition metals},
  author={Kresse, Georg and Hafner, JJPRB},
  journal={Physical Review B},
  volume={48},
  number={17},
  pages={13115},
  year={1993},
  publisher={APS}
}

@article{vasp2:kresse1996efficient,
  title={Efficient iterative schemes for ab initio total-energy calculations using a plane-wave basis set},
  author={Kresse, Georg and Furthm{\"u}ller, J{\"u}rgen},
  journal={Physical review B},
  volume={54},
  number={16},
  pages={11169},
  year={1996},
  publisher={APS}
}

@article{PBE:ernzerhof1999assessment,
  title={Assessment of the Perdew--Burke--Ernzerhof exchange-correlation functional},
  author={Ernzerhof, Matthias and Scuseria, Gustavo E},
  journal={The Journal of chemical physics},
  volume={110},
  number={11},
  pages={5029--5036},
  year={1999},
  publisher={American Institute of Physics}
}

@article{PAW:blochl1994projector,
  title={Projector augmented-wave method},
  author={Bl{\"o}chl, Peter E},
  journal={Physical review B},
  volume={50},
  number={24},
  pages={17953},
  year={1994},
  publisher={APS}
}

@article{HSE:heyd2003hybrid,
  title={Hybrid functionals based on a screened Coulomb potential},
  author={Heyd, Jochen and Scuseria, Gustavo E and Ernzerhof, Matthias},
  journal={The Journal of chemical physics},
  volume={118},
  number={18},
  pages={8207--8215},
  year={2003},
  publisher={American Institute of Physics}
}

@article{MD1:nose1984molecular,
  title={A molecular dynamics method for simulations in the canonical ensemble},
  author={Nos{\'e}, Sh{\=u}ichi},
  journal={Molecular physics},
  volume={52},
  number={2},
  pages={255--268},
  year={1984},
  publisher={Taylor \& Francis}
}

@article{MD2:hoover1985canonical,
  title={Canonical dynamics: Equilibrium phase-space distributions},
  author={Hoover, William G},
  journal={Physical review A},
  volume={31},
  number={3},
  pages={1695},
  year={1985},
  publisher={APS}
}

@article{DFPT:chaput2011phonon,
  title={Phonon-phonon interactions in transition metals},
  author={Chaput, Laurent and Togo, Atsushi and Tanaka, Isao and Hug, Gilles},
  journal={Physical Review B},
  volume={84},
  number={9},
  pages={094302},
  year={2011},
  publisher={APS}
}

@article{PHONOPY:togo2023implementation,
  title={Implementation strategies in phonopy and phono3py},
  author={Togo, Atsushi and Chaput, Laurent and Tadano, Terumasa and Tanaka, Isao},
  journal={Journal of Physics: Condensed Matter},
  year={2023},
  publisher={IOP Publishing}
}

@article{phonopy-phono3py-JPSJ,
  author  = {Togo, Atsushi},
  title   = {First-principles Phonon Calculations with Phonopy and Phono3py},
  journal = {J. Phys. Soc. Jpn.},
  volume  = {92},
  number  = {1},
  pages   = {012001},
  year    = {2023},
  doi     = {10.7566/JPSJ.92.012001}
}

@article{vaspkit:wang2021vaspkit,
  title={VASPKIT: A user-friendly interface facilitating high-throughput computing and analysis using VASP code},
  author={Wang, Vei and Xu, Nan and Liu, Jin-Cheng and Tang, Gang and Geng, Wen-Tong},
  journal={Computer Physics Communications},
  volume={267},
  pages={108033},
  year={2021},
  publisher={Elsevier}
}

@article{VESTA:momma2008vesta,
  title={VESTA: a three-dimensional visualization system for electronic and structural analysis},
  author={Momma, Koichi and Izumi, Fujio},
  journal={Journal of Applied crystallography},
  volume={41},
  number={3},
  pages={653--658},
  year={2008},
  publisher={International Union of Crystallography}
}

@article{fujishima1972electrochemical,
  title={Electrochemical photolysis of water at a semiconductor electrode},
  author={Fujishima, Akira and Honda, Kenichi},
  journal={nature},
  volume={238},
  number={5358},
  pages={37--38},
  year={1972},
  publisher={Nature Publishing Group UK London}
}

@article{guo2019fundamentals,
  title={Fundamentals of TiO2 photocatalysis: concepts, mechanisms, and challenges},
  author={Guo, Qing and Zhou, Chuanyao and Ma, Zhibo and Yang, Xueming},
  journal={Advanced Materials},
  volume={31},
  number={50},
  pages={1901997},
  year={2019},
  publisher={Wiley Online Library}
}

@article{nishioka2023photocatalytic,
  title={Photocatalytic water splitting},
  author={Nishioka, Shunta and Osterloh, Frank E and Wang, Xinchen and Mallouk, Thomas E and Maeda, Kazuhiko},
  journal={Nature Reviews Methods Primers},
  volume={3},
  number={1},
  pages={42},
  year={2023},
  publisher={Nature Publishing Group UK London}
}

@article{singla2021photocatalytic,
  title={Photocatalytic water splitting hydrogen production via environmental benign carbon based nanomaterials},
  author={Singla, Shelly and Sharma, Surbhi and Basu, Soumen and Shetti, Nagaraj P and Aminabhavi, Tejraj M},
  journal={International Journal of Hydrogen Energy},
  volume={46},
  number={68},
  pages={33696--33717},
  year={2021},
  publisher={Elsevier}
}

@article{baniasadi2013hybrid,
  title={Hybrid photocatalytic water splitting for an expanded range of the solar spectrum with cadmium sulfide and zinc sulfide catalysts},
  author={Baniasadi, E and Dincer, I and Naterer, GF},
  journal={Applied Catalysis A: General},
  volume={455},
  pages={25--31},
  year={2013},
  publisher={Elsevier}
}

@article{he2015novel,
  title={Novel-CdS-nanorod with stacking fault structures: Preparation and properties of visible-light-driven photocatalytic hydrogen production from water},
  author={He, Kai and Wang, Meng and Guo, Liejin},
  journal={Chemical Engineering Journal},
  volume={279},
  pages={747--756},
  year={2015},
  publisher={Elsevier}
}

@article{li2022challenges,
  title={Challenges of photocatalysis and their coping strategies},
  author={Li, Xinru and Chen, Yao and Tao, Ying and Shen, Li and Xu, Zhenmin and Bian, Zhenfeng and Li, Hexing},
  journal={Chem Catalysis},
  volume={2},
  number={6},
  pages={1315--1345},
  year={2022},
  publisher={Elsevier}
}

@article{bie2022challenges,
  title={Challenges for photocatalytic overall water splitting},
  author={Bie, Chuanbiao and Wang, Linxi and Yu, Jiaguo},
  journal={Chem},
  volume={8},
  number={6},
  pages={1567--1574},
  year={2022},
  publisher={Elsevier}
}

@article{kwon2015shape,
  title={Shape effects of cuprous oxide particles on stability in water and photocatalytic water splitting},
  author={Kwon, Yongwoo and Soon, Aloysius and Han, Haksoo and Lee, Hyunjoo},
  journal={Journal of Materials Chemistry A},
  volume={3},
  number={1},
  pages={156--162},
  year={2015},
  publisher={Royal Society of Chemistry}
}

@article{grozdanov1994electroless,
  title={Electroless chemical deposition technique for Cu2O thin films},
  author={Grozdanov, Ivan},
  journal={Materials Letters},
  volume={19},
  number={5-6},
  pages={281--285},
  year={1994},
  publisher={Elsevier}
}

@article{maeda2011photocatalytic,
  title={Photocatalytic water splitting using semiconductor particles: history and recent developments},
  author={Maeda, Kazuhiko},
  journal={Journal of Photochemistry and Photobiology C: Photochemistry Reviews},
  volume={12},
  number={4},
  pages={237--268},
  year={2011},
  publisher={Elsevier}
}

@article{maeda2010photocatalytic,
  title={Photocatalytic water splitting: recent progress and future challenges},
  author={Maeda, Kazuhiko and Domen, Kazunari},
  journal={The Journal of Physical Chemistry Letters},
  volume={1},
  number={18},
  pages={2655--2661},
  year={2010},
  publisher={ACS Publications}
}

@article{novoselov2004electric,
  title={Electric field effect in atomically thin carbon films},
  author={Novoselov, Kostya S and Geim, Andre K and Morozov, Sergei V and Jiang, De-eng and Zhang, Yanshui and Dubonos, Sergey V and Grigorieva, Irina V and Firsov, Alexandr A},
  journal={science},
  volume={306},
  number={5696},
  pages={666--669},
  year={2004},
  publisher={American Association for the Advancement of Science}
}

@article{yuan2023first,
  title={First-principles calculations of 2D Janus WSSiN2 monolayer for photocatalytic water splitting},
  author={Yuan, Gang and Ma, Xinguo and Liao, Jiajun and Xie, Tian and Xie, Qihai and Yuan, Zhongyong},
  journal={ACS Applied Nano Materials},
  volume={6},
  number={3},
  pages={1956--1964},
  year={2023},
  publisher={ACS Publications}
}

@article{novoselov20162d,
  title={2D materials and van der Waals heterostructures},
  author={Novoselov, Konstantin S and Mishchenko, Artem and Carvalho, Alexandra and Castro Neto, AH},
  journal={Science},
  volume={353},
  number={6298},
  pages={aac9439},
  year={2016},
  publisher={American Association for the Advancement of Science}
}

@article{chhowalla2013chemistry,
  title={The chemistry of two-dimensional layered transition metal dichalcogenide nanosheets},
  author={Chhowalla, Manish and Shin, Hyeon Suk and Eda, Goki and Li, Lain-Jong and Loh, Kian Ping and Zhang, Hua},
  journal={Nature chemistry},
  volume={5},
  number={4},
  pages={263--275},
  year={2013},
  publisher={Nature Publishing Group UK London}
}

@article{hisatomi2014recent,
  title={Recent advances in semiconductors for photocatalytic and photoelectrochemical water splitting},
  author={Hisatomi, Takashi and Kubota, Jun and Domen, Kazunari},
  journal={Chemical Society Reviews},
  volume={43},
  number={22},
  pages={7520--7535},
  year={2014},
  publisher={Royal Society of Chemistry}
}

@article{kang2021electronic,
  title={Electronic structure of aqueous two-dimensional photocatalyst},
  author={Kang, Dawei and Kong, Xianghua and Michaud-Rioux, Vincent and Chen, Ying-Chih and Mi, Zetian and Guo, Hong},
  journal={npj Computational Materials},
  volume={7},
  number={1},
  pages={47},
  year={2021},
  publisher={Nature Publishing Group UK London}
}

@article{jun2023two,
  title={Two-dimensional materials for photoelectrochemical water splitting},
  author={Jun, Sang Eon and Lee, Jae Kwan and Jang, Ho Won},
  journal={Energy Advances},
  volume={2},
  number={1},
  pages={34--53},
  year={2023},
  publisher={Royal Society of Chemistry}
}

@article{kudo2009heterogeneous,
  title={Heterogeneous photocatalyst materials for water splitting},
  author={Kudo, Akihiko and Miseki, Yugo},
  journal={Chemical Society Reviews},
  volume={38},
  number={1},
  pages={253--278},
  year={2009},
  publisher={Royal Society of Chemistry}
}

@article{low2017heterojunction,
  title={Heterojunction photocatalysts},
  author={Low, Jingxiang and Yu, Jiaguo and Jaroniec, Mietek and Wageh, Swelm and Al-Ghamdi, Ahmed A},
  journal={Advanced materials},
  volume={29},
  number={20},
  pages={1601694},
  year={2017},
  publisher={Wiley Online Library}
}

@article{zhang2012recent,
  title={Recent progress on graphene-based photocatalysts: current status and future perspectives},
  author={Zhang, Nan and Zhang, Yanhui and Xu, Yi-Jun},
  journal={Nanoscale},
  volume={4},
  number={19},
  pages={5792--5813},
  year={2012},
  publisher={Royal Society of Chemistry}
}

@article{liu2025two,
  title={Two-dimensional C20-based monolayers and heterostructures for photocatalytic overall water splitting},
  author={Liu, Miao and Yang, Chuan-Lu and Li, Xiaohu and Zhan, Li-Bo and Li, Yongqing},
  journal={npj 2D Materials and Applications},
  volume={9},
  number={1},
  pages={25},
  year={2025},
  publisher={Nature Publishing Group UK London}
}

@article{cheng2025tunable,
  title={Tunable Z-scheme photocatalytic activity in Sc2CBrX/MoTe2 (X= Br, Cl) heterostructures: a combined electronic and non-adiabatic dynamics study},
  author={Cheng, Yuan-Yuan and Li, Xiao-Ting and Yang, Chuan-Lu and Li, Xiaohu and Zhao, Wenkai and Liu, Yuliang and Gao, Feng},
  journal={Journal of Materials Chemistry A},
  volume={13},
  number={43},
  pages={37385--37395},
  year={2025},
  publisher={The Royal Society of Chemistry}
}

@article{zhou2023sth,
  title={Solar-to-hydrogen efficiency of more than 9\% in photocatalytic water splitting},
  author={Zhou, Peng and Navid, Ishtiaque Ahmed and Ma, Yongjin and Xiao, Yixin and Wang, Ping and Ye, Zhengwei and Zhou, Baowen and Sun, Kai and Mi, Zetian and others},
  journal={Nature},
  volume={613},
  pages={66--70},
  year={2023},
  doi={10.1038/s41586-022-05399-1}
}

@article{fu2024perovskite,
  title={Photocatalytic Overall Water Splitting with a Solar-to-Hydrogen Conversion Efficiency Exceeding 2\% through Halide Perovskite},
  author={Fu, Hui and Zhang, Qianqian and Liu, Yuanyuan and Zheng, Zhaoke and Cheng, Hefeng and Huang, Baibiao and Wang, Peng},
  journal={Angewandte Chemie International Edition},
  volume={63},
  pages={e202411016},
  year={2024},
  doi={10.1002/anie.202411016}
}

@article{gajdos2006linear,
  title={Linear optical properties in the projector-augmented wave methodology},
  author={Gajdo{\v{s}}, M. and Hummer, K. and Kresse, G. and Furthm{\"u}ller, J. and Bechstedt, F.},
  journal={Physical Review B},
  volume={73},
  number={4},
  pages={045112},
  year={2006},
  doi={10.1103/PhysRevB.73.045112}
}

@article{li2018twodimensional,
  title={Two-dimensional models for the optical response of thin films},
  author={Li, Yilei and Heinz, Tony F.},
  journal={2D Materials},
  volume={5},
  number={2},
  pages={025021},
  year={2018},
  doi={10.1088/2053-1583/aab0cf}
}

@article{grimme2010consistent,
  title={A consistent and accurate ab initio parametrization of density functional dispersion correction (DFT-D) for the 94 elements H-Pu},
  author={Grimme, Stefan and Antony, Jens and Ehrlich, Stephan and Krieg, Helge},
  journal={The Journal of Chemical Physics},
  volume={132},
  number={15},
  pages={154104},
  year={2010}
}

@misc{ASTM_G173,
  title        = {Standard Tables for Reference Solar Spectral Irradiances: Direct Normal and Hemispherical on 37 Degree Tilted Surface},
  organization = {ASTM International},
  number       = {ASTM G173-03(2020)},
  year         = {2020},
  doi          = {10.1520/G0173-03R20}
}

@article{majerus2023anisotropy,
  title={Anisotropy and effective medium approach in the optical response of two-dimensional material heterostructures},
  author={Maj{\'e}rus, Bruno and Guillaume, Emerick and Kockaert, Pascal and Henrard, Luc},
  journal={Physical Review B},
  volume={108},
  number={24},
  pages={245412},
  year={2023},
  publisher={American Physical Society},
  doi={10.1103/PhysRevB.108.245412}
}

@article{lobet2016perfect,
  title={Perfect electromagnetic absorption using graphene and epsilon-near-zero metamaterials},
  author={Lobet, Michael and Majerus, Bruno and Henrard, Luc and Lambin, Philippe},
  journal={Physical Review B},
  volume={93},
  number={23},
  pages={235424},
  year={2016},
  publisher={American Physical Society},
  doi={10.1103/PhysRevB.93.235424}
}

@article{Chaoui2025_InNPtSSe,
  author  = {Chaoui, Khawla and Hamidani, Ali and Zanat, Kamel},
  title   = {A novel Janus InN/PtSSe heterostructure for highly efficient overall water splitting via a third-generation Z-scheme: A first-principles study},
  journal = {International Journal of Hydrogen Energy},
  volume  = {174},
  pages   = {151360},
  year    = {2025},
  doi     = {10.1016/j.ijhydene.2025.151360}
}

@article{Chaoui2025_XGaPbP,
  author  = {Chaoui, Khawla and Elaggoune, Warda and Henrard, Luc and Zanat, Kamel and Achehboune, Mohamed},
  title   = {Novel Janus XGa--PbP (X = S, Se) monolayers: excellent photocatalysts for overall water splitting},
  journal = {RSC Advances},
  volume  = {15},
  number  = {38},
  pages   = {31076--31087},
  year    = {2025},
  doi     = {10.1039/D5RA04300J}
}

@article{Chaoui2024_GeSnXY,
  author  = {Chaoui, Khawla and Elaggoune, Warda and Zanat, Kamel},
  title   = {Understanding the potential of 2D Janus GeSnX$_2$ and its quaternary derivatives XGe--SnY (X, Y = P, As, X $\neq$ Y) for photocatalytic overall water splitting},
  journal = {International Journal of Hydrogen Energy},
  volume  = {91},
  pages   = {1221--1231},
  year    = {2024},
  doi     = {10.1016/j.ijhydene.2024.10.107}
}

@article{Chaoui2024_BAsGeC,
  author  = {Chaoui, Khawla and Zanat, Kamel and Elaggoune, Warda and Henrard, Luc and Achehboune, Mohamed},
  title   = {Theoretical insights into Z-scheme BAs/GeC van der Waals heterostructure for high-efficiency solar cell},
  journal = {RSC Advances},
  volume  = {14},
  pages   = {39625--39635},
  year    = {2024},
  doi     = {10.1039/D4RA08369E}
}

@article{Elaggoune2025_BX,
  author  = {Elaggoune, Warda and Abdullahi, Yusuf Zuntu and Chaoui, Khawla},
  title   = {Multifunctional BX (X = P, As, Sb) biphenylene-based derivatives as water splitting photocatalysts and toxic gas sensors: A first-principles study},
  journal = {International Journal of Hydrogen Energy},
  volume  = {156},
  pages   = {150099},
  year    = {2025},
  doi     = {10.1016/j.ijhydene.2025.150099}
}

@article{habli2025computational,
  title={Computational analysis of solar light harvesting properties of TiO2-BiVO4 inverse opals for applications in photocatalysis},
  author={Habli, Oumayma and Madanu, Thomas L and Su, Bao-Lian and Deparis, Olivier},
  journal={Journal of Energy Chemistry},
  year={2025},
  publisher={Elsevier}
}

@article{CdS:review2023,
  title={Review on CdS-derived photocatalysts for solar photocatalytic applications--Advances and challenges},
  author={Ahmad, Irshad and Ahmed, Samia Ben and Shabir, Maria and Imran, Muhammad and Hassan, Ahmed M and Alatawi, Naifa S},
  journal={Journal of Industrial and Engineering Chemistry},
  volume={130},
  pages={105--124},
  year={2024},
  publisher={Elsevier}
}

@article{10.1021/acsomega.0c02804,
    author = {Ashwin Kishore, M. R. and Larsson, Karin and Ravindran, Ponniah},
    title = {Two-Dimensional CdX/C2N (X = S, Se) Heterostructures
as Potential Photocatalysts for Water Splitting: A DFT Study},
    journal = {ACS Omega},
    volume = {5},
    number = {37},
    pages = {23762-23768},
    year = {2020},
    month = {09},
    issn = {2470-1343},
    doi = {10.1021/acsomega.0c02804},
    url = {https://doi.org/10.1021/acsomega.0c02804},
    eprint = {https://pubs.acs.org/acsodf/article-pdf/5/37/23762/67719795/acsomega.0c02804.pdf},
}

@Article{nano13182535,
AUTHOR = {El Hamdaoui, Jawad and Pérez, Laura M. and Ojeda-Martínez, Miguel and El Ouarie, Nassima and Díaz, Pablo and Laroze, David and Feddi, El Mustapha},
TITLE = {First Principle Study on the Effect of Strain on the Electronic Structure and Carrier Mobility of the Janus MoSTe and WSTe Monolayers},
JOURNAL = {Nanomaterials},
VOLUME = {13},
YEAR = {2023},
NUMBER = {18},
ARTICLE-NUMBER = {2535},
URL = {https://www.mdpi.com/2079-4991/13/18/2535},
PubMedID = {37764563},
ISSN = {2079-4991},
DOI = {10.3390/nano13182535}
}

@article{10.1039/c6nr02923j,
    author = {Xie, Meiqiu and Zhang, Shengli and Cai, Bo and Zhu, Zhen and Zou, Yousheng and Zeng, Haibo},
    title = {Two-dimensional BX (X = P, As, Sb) semiconductors with mobilities approaching graphene},
    journal = {Nanoscale},
    volume = {8},
    number = {27},
    pages = {13407-13413},
    year = {2016},
    month = {07},
    issn = {2040-3364},
    doi = {10.1039/c6nr02923j},
    url = {https://doi.org/10.1039/c6nr02923j},
    eprint = {https://pubs.rsc.org/nr/article-pdf/8/27/13407/4951870/c6nr02923j.pdf},
}
\end{document}


\maketitle

\section{Structure and stability}

\begin{figure}[htbp]
\centering
\includegraphics[width=0.95\textwidth]{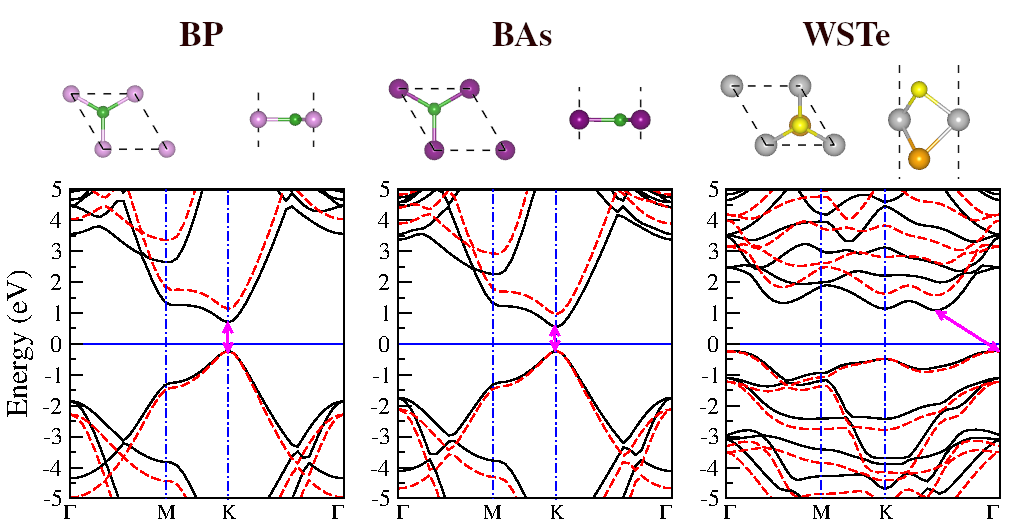}
\caption{Relaxed structures and calculated electronic band structures of the isolated BP, BAs, and Janus WSTe monolayers using HSE06.}
\label{fig:mono}
\end{figure}

Before constructing the heterostructures, we examined the structural and electronic properties of the isolated BP, BAs, and Janus WSTe monolayers. BP and BAs adopt planar hexagonal lattices, whereas WSTe consists of a W atomic plane between inequivalent S and Te planes. This out-of-plane compositional asymmetry breaks mirror symmetry and produces two chemically distinct surfaces. The relaxed structures and band structures of the monolayers are shown in Fig.~\ref{fig:mono}; their lattice constants and band gaps are compared with previously reported theoretical values in Table~\ref{tab:mono} \cite{BX:csahin2009monolayer,BX:islam2022tuning,BX:khurami2023strain,WSTe:patel2020high}.

The calculated lattice constants of BP, BAs, and WSTe are 3.21, 3.39, and 3.34~\AA, respectively. Using the WSTe lattice constant as a reference gives lattice mismatches of approximately 3.9\% for BP/WSTe and 1.5\% for BAs/WSTe. These relatively small mismatches permit the construction of commensurate heterostructure models, although the resulting in-plane strain should be taken into account when interpreting their properties.

\begin{table}[htbp]
\centering
\caption{Comparison of the calculated lattice constants
($a$) and HSE06 band gaps ($E_g$) of the isolated
BP, BAs, and Janus WSTe monolayers with previously
reported theoretical results.}
\label{tab:mono}
\begin{tabular}{lcccc}
\toprule
Material &
\multicolumn{2}{c}{$a$ (\AA)} &
\multicolumn{2}{c}{$E_g$ (eV)} \\
\cmidrule(lr){2-3}
\cmidrule(lr){4-5}
& This work & Literature
& This work & Literature \\
\midrule
BP   & 3.21 & 3.21 \cite{10.1039/c6nr02923j}
     & 1.36 & 1.36 \cite{10.1039/c6nr02923j} \\
BAs  & 3.39 & 3.39 \cite{10.1039/c6nr02923j}
     & 1.18 & 1.14 \cite{10.1039/c6nr02923j} \\
WSTe & 3.34 & 3.359 \cite{nano13182535}
     & 1.59 & 1.56 \cite{nano13182535} \\
\bottomrule
\end{tabular}
\end{table}

\begin{figure}[htbp]
\centering
\includegraphics[width=0.6\textwidth]{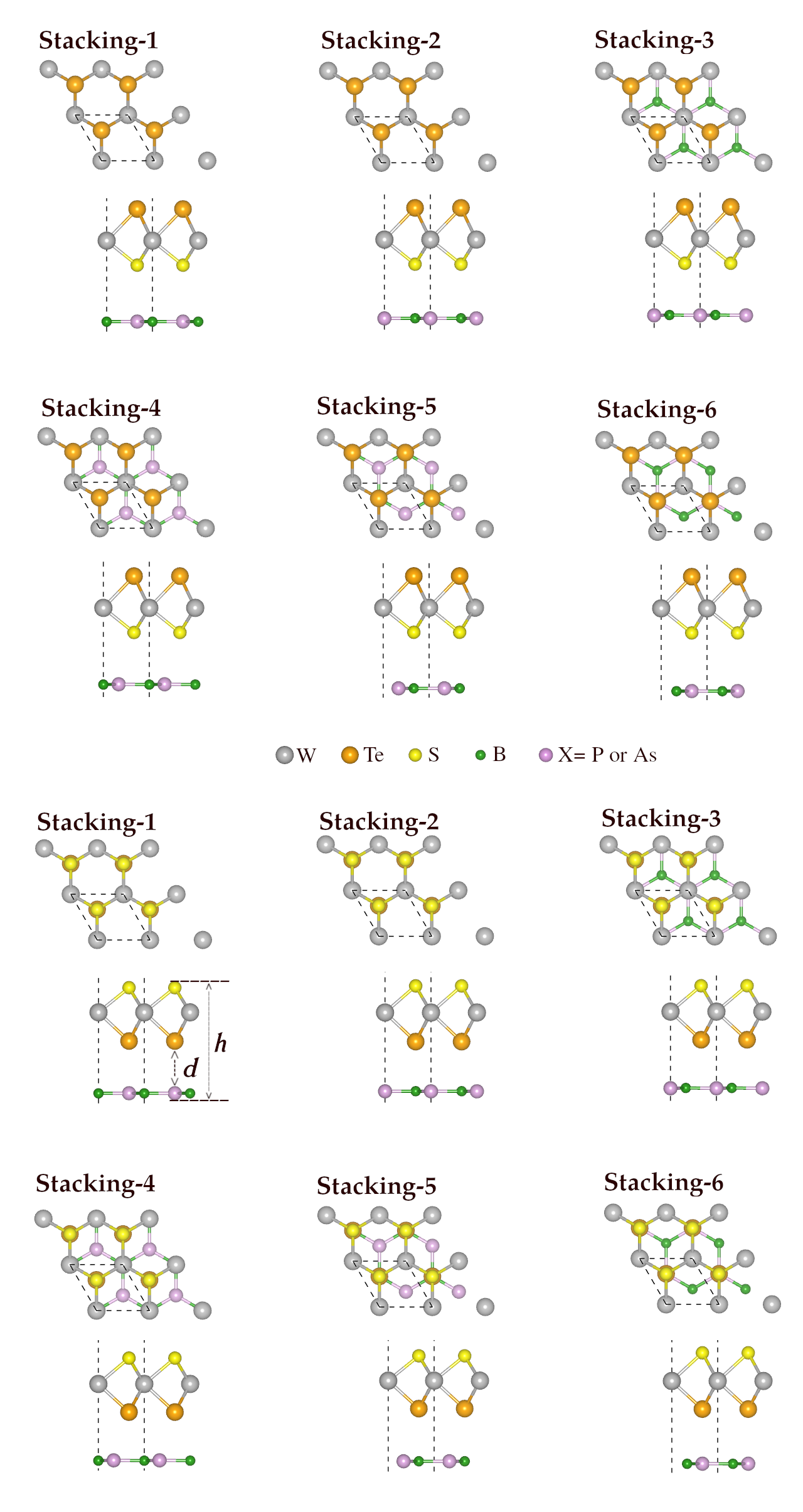}
\caption{Top and side views of the six stacking configurations of BX/WSTe ($X$ = P, As), for S- and Te-facing interfaces. The interlayer distance $d$ is indicated.}
\label{fig:stackings}
\end{figure}

To examine the effect of interfacial registry, six stacking configurations were considered for each BX/WSTe ($X$ = P, As) heterostructure and for both S- and Te-facing orientations of WSTe. The S-facing systems are labeled BP/SWTe and BAs/SWTe, while the Te-facing systems are labeled BP/TeWS and BAs/TeWS. The considered configurations are shown in Fig.~\ref{fig:stackings}. Stackings 1 and 2 are AA-type arrangements: in Stacking-1, the interfacial chalcogen atom is aligned above X and W above B; in Stacking-2, the interfacial chalcogen atom is aligned above B and W above X. Stackings 3--6 correspond to laterally shifted registries. In Stackings 3 and 4, W is aligned above X and B, respectively, while the interfacial chalcogen occupies a hollow site. In Stackings 5 and 6, W occupies a hollow site and the interfacial chalcogen is aligned above B and X, respectively.

The binding energy of each relaxed heterostructure was evaluated as
\begin{equation}
E_b=E_{\mathrm{hetero}}-\left(E_{\mathrm{BX}}+E_{\mathrm{WSTe}}\right),
\label{eq:binding}
\end{equation}
where $E_{\mathrm{hetero}}$ is the total energy of the heterostructure and $E_{\mathrm{BX}}$ and $E_{\mathrm{WSTe}}$ refer to the isolated constituent monolayers calculated with consistent settings. For a quantitative binding-energy comparison, the isolated-layer reference calculations were performed using the same in-plane cell and strain state as the corresponding heterostructure. The negative calculated values indicate that the assembled configurations are energetically favored relative to the specified isolated-layer reference states.

\begin{table}[htbp]
\centering
\caption{Optimized in-plane lattice parameter ($a$), binding energy ($E_b$), and interlayer separation ($d$) for six registries of the S- and Te-facing BX/WSTe heterostructures. Binding energies are reported per simulation cell.}
\label{tab:structural}
\begin{tabular}{ccccccc}
\toprule
& \multicolumn{3}{c}{BP/WSTe} & \multicolumn{3}{c}{BAs/WSTe} \\
\cmidrule(lr){2-4}\cmidrule(lr){5-7}
Stacking & $a$ (\AA) & $E_b$ (eV) & $d$ (\AA) & $a$ (\AA) & $E_b$ (eV) & $d$ (\AA) \\
\midrule
\multicolumn{7}{c}{\textbf{S-facing interface}} \\
\midrule
1 & 3.259 & -0.829 & 3.62 & 3.354 & -0.908 & 3.49 \\
2 & 3.259 & -0.838 & 3.53 & 3.354 & -0.915 & 3.41 \\
3 & 3.260 & -0.852 & 3.36 & 3.356 & -0.929 & 3.39 \\
4 & 3.259 & \textbf{-0.892} & 3.20 & 3.354 & -0.963 & 3.22 \\
5 & 3.258 & -0.863 & 3.23 & 3.353 & -0.961 & 3.26 \\
6 & 3.259 & -0.819 & 3.64 & 3.354 & -0.898 & 3.51 \\
\midrule
\multicolumn{7}{c}{\textbf{Te-facing interface}} \\
\midrule
1 & 3.257 & -0.840 & 3.60 & 3.353 & -0.915 & 3.63 \\
2 & 3.258 & -0.847 & 3.58 & 3.354 & -0.918 & 3.62 \\
3 & 3.259 & -0.872 & 3.50 & 3.355 & -0.942 & 3.56 \\
4 & 3.257 & \textbf{-0.930 }& 3.41 & 3.352 & -0.998 & 3.44 \\
5 & 3.257 & -0.888 & 3.45 & 3.352 & -0.986 & 3.48 \\
6 & 3.257 & -0.832 & 3.59 & 3.352 & -0.903 & 3.65 \\
\bottomrule
\end{tabular}
\end{table}

As shown in Table~\ref{tab:structural}, Stacking-4 is the lowest-energy registry among the six configurations tested for each system. Its binding energies are $-0.892$, $-0.963$, $-0.930$, and $-0.998$~eV for BP/SWTe, BAs/SWTe, BP/TeWS, and BAs/TeWS, respectively; the corresponding interlayer separations are 3.20, 3.22, 3.41, and 3.44~\AA. Stacking-5 lies only 0.029~eV above Stacking-4 for BP/SWTe and 0.002~eV above it for BAs/SWTe, showing that the relative preference can be small. These energies and separations are consistent with predominantly van der Waals interlayer coupling. Stacking-4 was therefore selected for the subsequent calculations, while its near-degeneracy with Stacking-5 should be kept in mind.

\subsection{Dynamical and short-time thermal stability}

\begin{figure}[htbp]
\centering
\includegraphics[width=\textwidth]{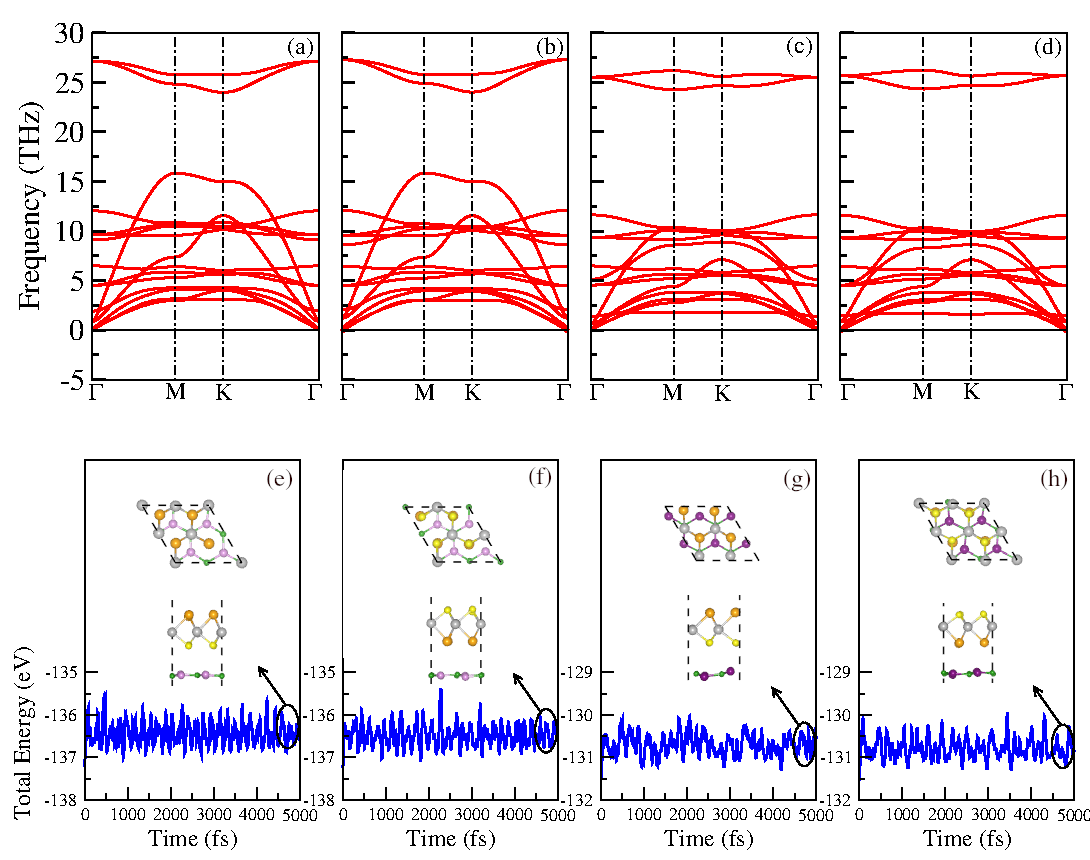}
\caption{Stability analysis of the four Stacking-4 heterostructures: (a--d) calculated phonon dispersions and (e--h) AIMD total-energy trajectories at 300~K for 5~ps.}
\label{fig:stability}
\end{figure}

The stability of the relaxed Stacking-4 heterostructures was examined using phonon dispersions and \textit{ab initio} molecular dynamics (AIMD). Figure~\ref{fig:stability}(a--d) shows the phonon dispersions; no imaginary branches are apparent in the calculated spectra, supporting dynamical stability at the adopted computational level. Figure~\ref{fig:stability}(e--h) shows AIMD trajectories at 300~K over 5~ps. The energies fluctuate without obvious structural degradation during the simulated time interval. These results support short-time structural stability at room temperature, but do not establish long-term stability or chemical stability in an aqueous photocatalytic environment.

\subsection{In-plane elastic properties}

In-plane mechanical stability was assessed using the elastic constants $C_{11}$ and $C_{12}$ obtained from the stress--strain approach. For an in-plane hexagonal elastic response \cite{stress:born1996dynamical,mechstab:wang2022high}, $C_{22}=C_{11}$ and $C_{66}=(C_{11}-C_{12})/2$. The relevant positivity conditions can be written as $C_{11}>|C_{12}|$, or equivalently $C_{11}-C_{12}>0$ and $C_{11}+C_{12}>0$. The two-dimensional Young's and shear moduli were calculated as
\begin{align}
Y_{2\mathrm{D}} &= \frac{C_{11}^{2}-C_{12}^{2}}{C_{11}}, \\
G_{2\mathrm{D}} &= \frac{C_{11}-C_{12}}{2}.
\end{align}
All quantities in Table~\ref{tab:elastic_constants} are expressed in N/m. The derived moduli were calculated from the reported elastic constants and rounded to two decimal places.

\begin{table}[htbp]
\centering
\caption{In-plane elastic constants and derived two-dimensional Young's ($Y_{2\mathrm{D}}$) and shear ($G_{2\mathrm{D}}$) moduli. All values are in N/m.}
\label{tab:elastic_constants}
\begin{tabular}{lrrrr}
\toprule
Material & $C_{11}$ & $C_{12}$ & $Y_{2\mathrm{D}}$ & $G_{2\mathrm{D}}$ \\
\midrule
BP        & 145.90 & 38.62 & 135.68 & 53.64 \\
BAs       & 124.94 & 34.43 & 115.45 & 45.26 \\
WSTe      & 124.91 & 19.11 & 121.99 & 52.90 \\
BAs/SWTe  & 249.04 & 60.13 & 234.52 & 94.46 \\
BAs/TeWS  & 252.77 & 59.81 & 238.62 & 96.48 \\
BP/SWTe   & 278.91 & 51.26 & 269.49 & 113.83 \\
BP/TeWS   & 278.88 & 51.42 & 269.40 & 113.73 \\
\bottomrule
\end{tabular}
\end{table}

The calculated elastic constants satisfy the in-plane stability conditions for all four heterostructures. In the BP-based systems, $C_{11}$ is approximately 279~N/m and $Y_{2\mathrm{D}}$ is approximately 269~N/m; the corresponding values for the BAs-based systems are approximately 249--253 and 235--239~N/m, respectively. The heterostructure moduli exceed the values of the individual monolayers, as expected when two layers contribute to the effective in-plane response. This increase should not, on its own, be interpreted as evidence of strong interlayer bonding. It supports resistance to small in-plane deformations within the considered elastic model.

\section{Effect of pH on the solar-to-hydrogen efficiency}

To examine the role of pH in the predicted photocatalytic performance, the hydrogen- and oxygen-evolution driving forces, $\chi_{H_2}$ and $\chi_{O_2}$, were evaluated using the pH-dependent redox energies specified in the main manuscript. At each pH, the minimum photon-energy threshold $E$ was determined using the adopted Fu-type prescription. Solar-to-hydrogen (STH) efficiencies are reported using both an idealized full-absorption treatment, $\eta_{\mathrm{STH}}^{\mathrm{Fu}}$, and the calculation weighted by the in-plane single-sheet absorptance $A_{\parallel}(\hbar\omega)$, $\eta_{\mathrm{STH}}^{A}$.

\begin{table}[htbp]
\centering
\caption{Calculated pH-dependent water-redox driving forces,
photon-energy threshold, and theoretical STH efficiencies
for BP/TeWS under the ideal full-absorption assumption
and using the calculated absorptance, before and after
the surface-potential difference correction.}
\label{tab:BPTeWS_pH}
\resizebox{\linewidth}{!}{%
\begin{tabular}{rrrrrrrr}
\toprule
pH & $\chi_{H_2}$ (eV) & $\chi_{O_2}$ (eV)
& $E$ (eV)
& $\eta_{\mathrm{STH}}^{\mathrm{Fu}}$ (\%)
& $\eta_{\mathrm{STH}}^{\prime,\mathrm{Fu}}$ (\%)
& $\eta_{\mathrm{STH}}^{A}$ (\%)
& $\eta_{\mathrm{STH}}^{\prime,A}$ (\%) \\
\midrule
0.5 & 0.361 & 0.070 & 0.981 & 60.569 & 33.631 & 2.957 & 2.870 \\
1.0 & 0.331 & 0.099 & 0.951 & 62.593 & 34.755 & 2.977 & 2.889 \\
1.5 & 0.301 & 0.129 & 0.921 & 63.994 & 35.533 & 2.991 & 2.902 \\
2.0 & 0.272 & 0.158 & 0.892 & 64.012 & 35.543 & 2.991 & 2.902 \\
2.5 & 0.243 & 0.188 & 0.863 & 64.090 & 35.586 & 2.992 & 2.903 \\
3.0 & 0.213 & 0.217 & 0.833 & 64.683 & 35.915 & 2.997 & 2.908 \\
3.5 & 0.184 & 0.246 & 0.820 & 65.432 & 36.331 & 3.003 & 2.913 \\
4.0 & 0.154 & 0.276 & 0.820 & 65.432 & 36.331 & 3.003 & 2.913 \\
4.5 & 0.125 & 0.305 & 0.820 & 65.432 & 36.331 & 3.003 & 2.913 \\
5.0 & 0.095 & 0.335 & 0.820 & 65.432 & 36.331 & 3.003 & 2.913 \\
5.5 & 0.066 & 0.364 & 0.820 & 65.432 & 36.331 & 3.003 & 2.913 \\
6.0 & 0.036 & 0.394 & 0.820 & 65.432 & 36.331 & 3.003 & 2.913 \\
6.5 & 0.007 & 0.423 & 0.820 & 65.432 & 36.331 & 3.003 & 2.913 \\
\bottomrule
\end{tabular}%
}
\end{table}

Within the band-edge criterion adopted in the main manuscript, BP/TeWS is predicted to be thermodynamically capable of overall water splitting for approximately ($0 < \mathrm{pH} < 6.6$). After applying the surface-potential correction, the ideal full-absorption STH efficiency increases from 33.631\% to 36.331\%, whereas the corresponding absorptance-weighted efficiency ranges from 2.870\% to 2.913\%.

\begin{table}[htbp]
\centering
\caption{Calculated pH-dependent water-redox driving forces,
photon-energy threshold, and theoretical STH efficiencies
for BAs/TeWS, evaluated under the ideal full-absorption
assumption and using the calculated absorptance,
before and after the surface-potential correction.}
\label{tab:BASTeWS_pH}
\small
\resizebox{\linewidth}{!}{%
\begin{tabular}{rrrrrrrr}
\toprule
pH & $\chi_{H_2}$ (eV) & $\chi_{O_2}$ (eV)
& $E$ (eV)
& $\eta_{\mathrm{STH}}^{\mathrm{Fu}}$ (\%)
& $\eta_{\mathrm{STH}}^{\prime,\mathrm{Fu}}$ (\%)
& $\eta_{\mathrm{STH}}^{A}$ (\%)
& $\eta_{\mathrm{STH}}^{\prime,A}$ (\%) \\
\midrule
4.0 & 0.244 & 0.016 & 0.964 & 61.726 & 33.811 & 3.288 & 3.177 \\
4.5 & 0.215 & 0.046 & 0.935 & 63.566 & 34.819 & 3.307 & 3.195 \\
5.0 & 0.185 & 0.075 & 0.920 & 64.002 & 35.058 & 3.311 & 3.199 \\
5.5 & 0.156 & 0.105 & 0.920 & 64.002 & 35.058 & 3.311 & 3.199 \\
6.0 & 0.126 & 0.134 & 0.920 & 64.002 & 35.058 & 3.311 & 3.199 \\
6.5 & 0.097 & 0.164 & 0.920 & 64.002 & 35.058 & 3.311 & 3.199 \\
7.0 & 0.067 & 0.193 & 0.920 & 64.002 & 35.058 & 3.311 & 3.199 \\
7.5 & 0.038 & 0.223 & 0.920 & 64.002 & 35.058 & 3.311 & 3.199 \\
8.0 & 0.008 & 0.252 & 0.920 & 64.002 & 35.058 & 3.311 & 3.199 \\
\bottomrule
\end{tabular}%
}
\end{table}

Similarly, BAs/TeWS is predicted to support overall water splitting within ($3.7 < \mathrm{pH} < 8.1$). The surface-potential-corrected ideal STH efficiency varies from 33.811\% to approximately 35.058\%, while the corrected absorptance-weighted efficiency ranges from 3.177\% to 3.199\%. These theoretical estimates do not account explicitly for reaction kinetics, charge-carrier recombination, or aqueous stability.

\bibliographystyle{unsrt}
\bibliography{rsc}